\documentclass{article}

\PassOptionsToPackage{numbers}{natbib}
\usepackage[preprint]{neurips_2026}

\usepackage[utf8]{inputenc} 
\usepackage[T1]{fontenc}    
\usepackage{hyperref}       
\usepackage{url}            
\usepackage{booktabs}       
\usepackage{amsmath}        
\usepackage{amsfonts}       
\usepackage{nicefrac}       
\usepackage{microtype}      
\usepackage{xcolor}         
\usepackage{graphicx}       
\usepackage{colortbl}       
\usepackage{wrapfig}        

\usepackage{graphicx}
\usepackage{amssymb}
\usepackage{multirow}
\usepackage{stfloats}
\usepackage[table]{xcolor}
\usepackage{adjustbox}

\usepackage{algorithm}
\usepackage{algorithmic}

\title{Geometric Gain Graph: Zero-Token Graph Construction for Multi-Hop RAG}

\author{%
  Zeliang Li\textsuperscript{1}\quad
  Xiaofen Xing\textsuperscript{1}\thanks{Corresponding authors: Xiaofen Xing and Xiangmin Xu.}\quad
  Kailing Guo\textsuperscript{1}\quad
  Wenyu Tao\textsuperscript{1}\quad
  Xiangmin Xu\textsuperscript{2,1}\footnotemark[1]\\
  \textsuperscript{1}South China University of Technology\\
  \textsuperscript{2}Foshan University\\
  {\small\texttt{zeliang0li@163.com}\quad\texttt{xfxing@scut.edu.cn}\quad\texttt{guokl@scut.edu.cn}}\\
  {\small\texttt{eetaowenyu@mail.scut.edu.cn}\quad\texttt{xmxu@scut.edu.cn}}
}

\begin{document}

\maketitle

\begin{abstract}
Existing Retrieval-Augmented Generation (RAG) systems struggle with multi-hop reasoning due to a fundamental inability to balance relevance and novelty. Traditional dense retrievers are constrained by surface-level matching, falling into a homogenized "Similarity Trap". Conversely, emerging Graph RAGs attempt to explore novel documents via heuristic entity extraction, but they introduce prohibitive Large Language Model (LLM) overhead and easily drift into off-topic "Novelty Traps". To fundamentally resolve this contradiction, we propose Geometric Gain Graph (G$^3$RAG), the novel zero-token graph construction paradigm for RAG. G$^3$RAG completely discards the expensive heuristic extraction by LLMs, aiming instead to directly quantify the information gain between nodes through the native document feature space. To this end, we designed a concise edge weight criterion, $\cos\theta \cdot \sin\theta$, which naturally maps the trade-off between relevance and novelty into a geometric measurement of directional consistency and orthogonality in the vector space, thereby driving the token-free topological construction of the graph. Furthermore, we introduce a topological penalty to suppress the excessive connectivity of high-frequency hubs, forcing the transient diffusion process toward long-tail peripheral nodes that carry critical indirect evidence. Extensive experiments demonstrate that, while entirely eliminating LLM token costs during graph construction, G$^3$RAG significantly outperforms state-of-the-art Graph RAG baselines on complex multi-hop QA benchmarks. The code is available at: \url{https://github.com/deeplearning-LI/G3RAG}
\end{abstract}

\section{Introduction}
\label{sec:intro}
Retrieval-Augmented Generation (RAG) \cite{lewis2020retrieval} mitigates LLM hallucinations via external knowledge, emerging as a core paradigm for modern generative AI \cite{li-etal-2025-unilr,tan-etal-2025-prospect}. However, when confronted with multi-hop query scenarios, traditional dense retrieval-based RAG mechanisms are prone to falling into the "\textbf{Similarity Trap}". Specifically, relying exclusively on surface-level semantic matching tends to retrieve highly homogenous documents \cite{lee2024hybgrag,peng2024graphretrievalaugmentedgenerationsurvey}. This low variance hinders the acquisition of novel, critical supporting evidence, thereby degrading the final generation quality.

\begin{figure*}
    \centering
    \includegraphics[width=1.0\textwidth]{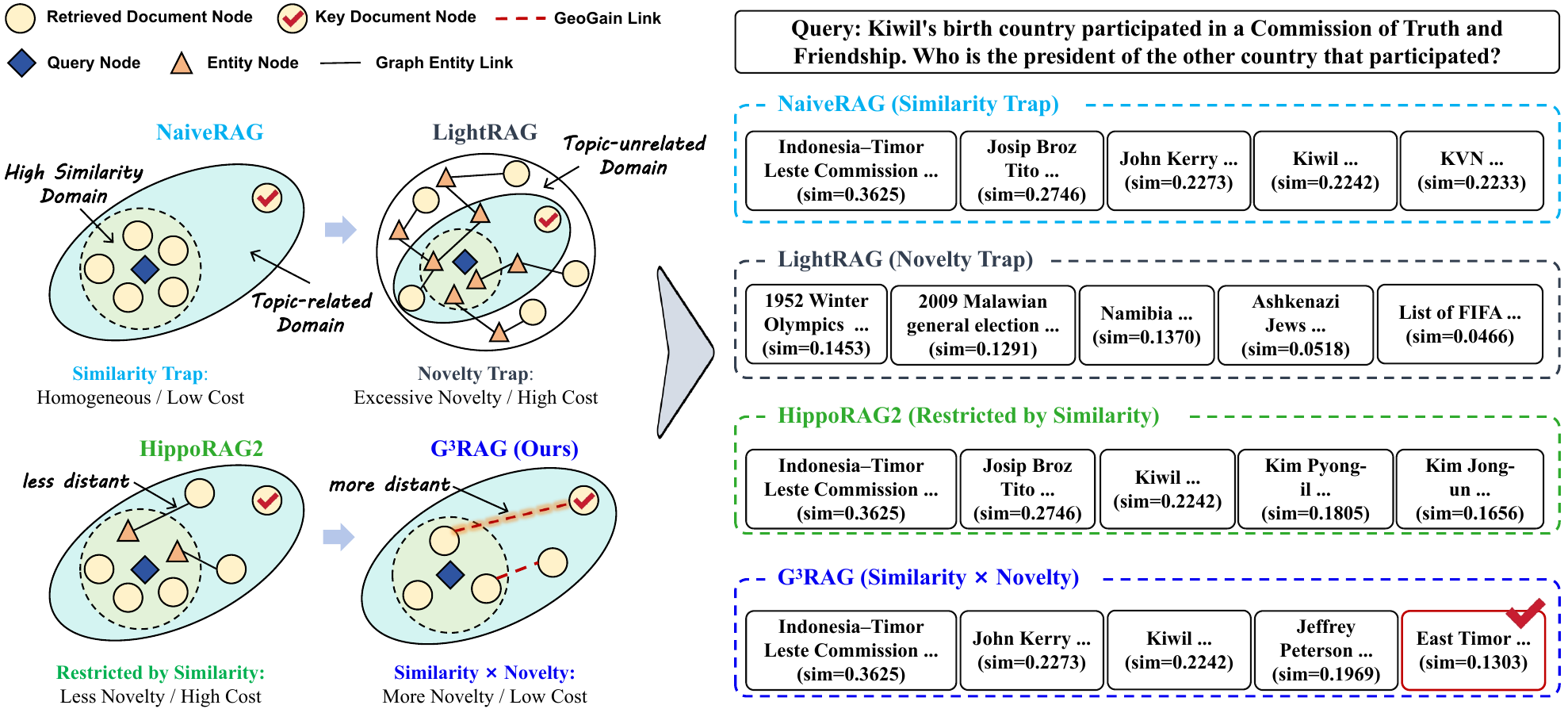}
    \caption{Illustration of RAG challenges. (a) Relying solely on similarity, NaiveRAG falls into a homogenized \textbf{Similarity Trap}. Conversely, LightRAG overemphasizes relational connections, retrieving off-topic documents and entering a \textbf{Novelty Trap}. Constrained by surface similarity, HippoRAG2 lacks sufficient novelty. In contrast, G$^3$RAG maintains adequate novelty through geometric gain while leveraging similarity. (b) Displays the similarity between retrieved documents and the query. NaiveRAG scores excessively high and LightRAG excessively low. Dominated by similarity, HippoRAG2 cannot efficiently discover more distant key documents like G$^3$RAG.}
    \label{fig:challenge}
\end{figure*}

Recently, graph-based RAG approaches (e.g., Graph RAG \cite{edge2024local} and LightRAG \cite{guo2024lightrag}) have leveraged the robust comprehension capabilities of LLMs for heuristic entity extraction and association. While enabling access to non-homogenous documents, heuristic graph construction requires massive offline LLM invocations, causing prohibitive costs. Despite this expense, performance gains over advanced dense retrievers remain marginal, as unrestricted node connections inevitably retrieve off-topic corpora, trapping the system in a \textbf{"Novelty Trap"}. To mitigate this unconstrained expansion, advanced studies such as HippoRAG2 \cite{gutierrez2025rag} and LinearRAG \cite{zhuang2025linearrag} introduced document nodes alongside entity nodes to ground the retrieval process. However, this structural constraint inadvertently diminishes the capacity for novelty exploration. Because the retrieval process becomes inherently bound by the surface-level similarity of these document nodes, the system fails to bridge the gap to distant key documents required for multi-hop reasoning, inherently limiting its capacity for novelty exploration, as shown in Figure \ref{fig:challenge}. Consequently, deploying Graph RAG to resolve multi-hop queries continues to face significant challenges:
(1) \textbf{Lack of actively novelty modeling}: Existing methods remain constrained by similarity metrics and lack a rigorous mechanism to define and balance the trade-off between relevance and novelty.
(2) \textbf{Prohibitive computational overhead}: Graph construction and maintenance rely heavily on token-expensive LLMs for heuristic information extraction.

\begin{figure*}
    \centering
    \includegraphics[width=1.00\textwidth]{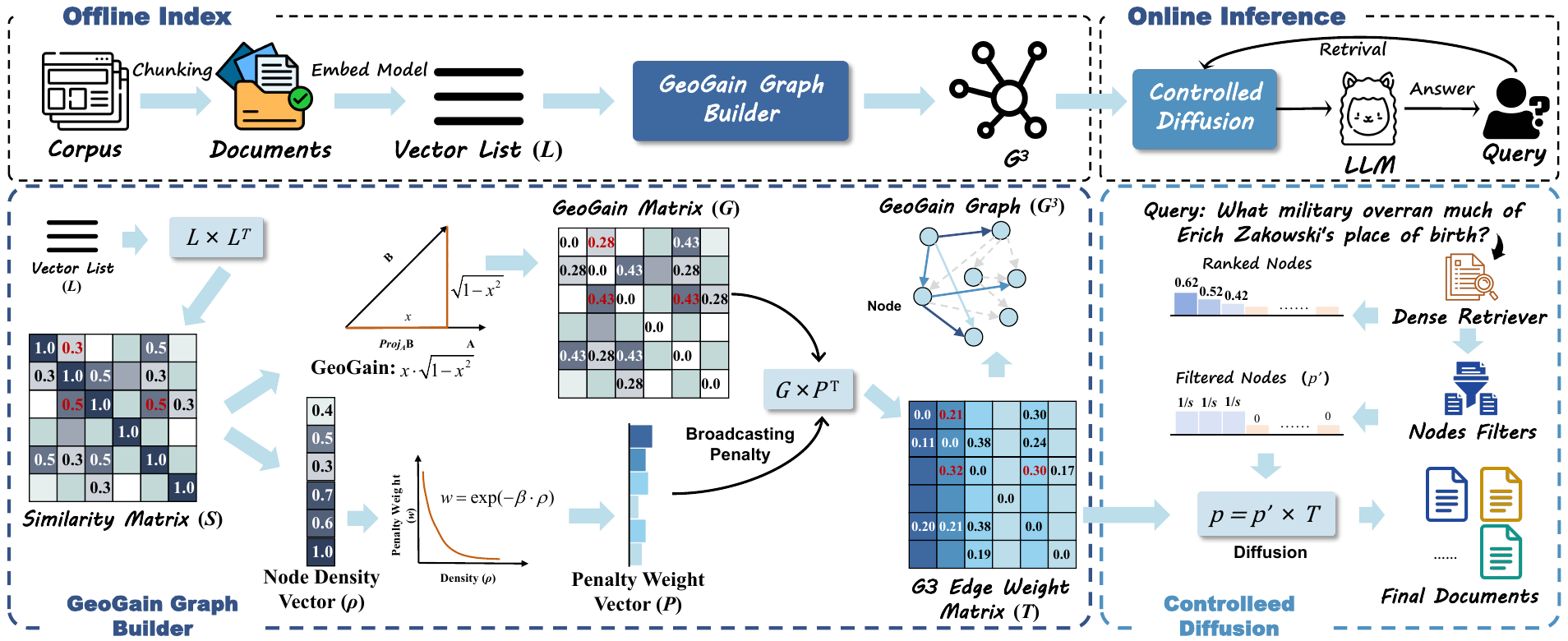}
    \caption{The overall framework of G$^3$RAG. It comprises two decoupled stages: Offline Geometric Graph Index and Online Controlled Diffusion Inference. Offline, the graph is constructed by defining edge weights as $\cos\theta \cdot \sin\theta$ to balance relevance and novelty, alongside a topological penalty to decay weights in prominently dense regions. During inference, G$^3$RAG first anchors high-confidence seed nodes via dense retrieval and LLM filtering, and then triggers a controlled transient diffusion process along the optimized edges to retrieve optimal multi-hop evidence.}
    \label{fig:framework}
\end{figure*}

To fundamentally address the aforementioned challenges, we propose a disruptive \textbf{G}raph RAG paradigm based on \textbf{G}eometric \textbf{G}ain, named G$^3$RAG as shown in Figure \ref{fig:framework}. Unlike entity-dependent methods, G$^3$RAG builds robust graphs directly from the geometric properties of document representations. This eliminates the need for extra neural encoder/decoders and achieves "\textbf{zero token}" consumption, thereby completely resolving the challenges of high construction and maintenance costs in traditional graph building. The core motivation of G$^3$RAG is founded upon an intuitive yet rigorous geometric assumption: balancing relevance and novelty in multi-hop retrieval essentially equates to seeking the optimal solution between directional consistency and orthogonality within the vector space. Specifically, the relevance between two documents can be characterized by the directional consistency of their vector representations (i.e., $\cos\theta \to 1$), while the introduced novelty (information gain) can be characterized by their orthogonality (i.e., $\sin\theta \to 1$). Based on this, G$^3$RAG ingeniously defines the edge weight between documents as the product of relevance and novelty ($\cos\theta \cdot \sin\theta$). This simple yet elegant geometric exploration naturally provides a LLM-free connection criterion for graph construction. Furthermore, we also introduce a topological penalty to penalize high-frequency nodes, compelling the graph to connect with rarer, peripheral nodes in order to preserve novelty. Inspired by the human cognitive mechanism of anchor-driven association \cite{el2024cellular}, we obtain the seed nodes of the graph through matching and filtering, and achieve the integration of optimal documents via controlled transient diffusion (association). \textbf{In summary}, the main contributions of this paper are as follows: (1) We propose G$^3$RAG, which achieves \textbf{zero-token} graph construction by modeling the geometric relationships of document representations. This is a pioneering and groundbreaking work in the RAG field. (2) To balance relevance and novelty, we design an edge-weighting criterion based on representation consistency and orthogonality. Coupled with a topological penalty to suppress high-frequency hubs, this mechanism guides the transient diffusion process to explore long-tail peripheral nodes for critical evidence. (3) Extensive experiments across challenging multi-hop QA benchmarks demonstrate that G$^3$RAG achieves SOTA performance. Notably, it significantly enhances complex reasoning and novelty exploration while strictly maintaining its zero-token graph construction advantage.

\section{Related Work}
\label{sec:relatedwork}
\subsection{Retrieval-Augmented Generation}
Retrieval-Augmented Generation (RAG) effectively mitigates LLM hallucinations but struggles to retrieve key documents for multi-hop queries \cite{zhu2025mitigating, zhuang2024efficientrag, shen2025gear}. Traditional Naive RAG relies on surface features, missing deep semantic correlations, while custom multi-hop models often lack generalization. Furthermore, optimization strategies like reranking and filtering fail to address the underlying flattened data structure \cite{glass2022re2g,wang2024corag,zhao2024meta}. Consequently, the retrieval process falls into a "similarity trap," unable to bridge semantic gaps and uncover latent documents essential for complex reasoning. Classical diversification methods explicitly account for novelty. Maximal Marginal Relevance (MMR) \cite{carbonell1998use} greedily balances query relevance against redundancy with previously selected documents, while determinantal point processes (DPPs) \cite{kulesza2012determinantal} use a kernel to jointly model item quality and subset diversity. When used to rerank a fixed candidate pool, both are constrained by its coverage and cannot recover evidence excluded during initial retrieval. Moreover, reducing similarity among selected documents does not by itself ensure complementary evidence for a multi-hop reasoning chain. These limitations motivate modeling relevance-conditioned novelty during graph construction and using the resulting links to expand retrieval beyond the initial candidates.

\subsection{Graph-based RAG}
To overcome the limitations of flattened data structures, RAG research has shifted towards Graph RAG, leveraging graph topology for knowledge retrieval \cite{zhang2025survey,guo2025empowering,li2025t}. For instance, methods like GraphRAG \cite{edge2024local} and LightRAG \cite{guo2024lightrag} use LLMs to extract entities and build complex connections for multi-hop queries. However, due to inherent extraction hallucinations and unverified relationships, these approaches often introduce redundant noise and fall into a "novelty trap," severely occupying the context window and sometimes underperforming advanced dense retrievers \cite{han2025rag,you2025ms,zhang2025qwen3,lee2024nv}. Recent works like HippoRAG2 \cite{gutierrez2025rag} and LinearRAG \cite{zhuang2025linearrag} address this by integrating document nodes into the graph architecture to constrain chaotic retrieval expansion. Nevertheless, because these methods fail to actively model novelty and relevance, they remain dominated by surface-level semantic similarity, leading to insufficient exploration of deep, novel, and critical evidence. In contrast, G$^3$RAG models both similarity and novelty through the geometric relationships between document nodes. By introducing geometric prior-based edge weight constraints and controlled transient diffusion, G$^3$RAG achieves \textbf{zero token} graph construction. This allows the architecture to robustly expand semantic boundaries while maintaining baseline relevance.

\section{Problem Formulation}
\label{sec:problem}
Graph RAG retrieves external evidence via a structured knowledge graph $\mathcal{G}$ to enhance LLMs' multi-hop reasoning. For a query $q$, the system retrieves an optimal evidence subset $\mathcal{D}_{ret}$ using $\text{Retrieve}(\cdot)$, and the language model $\mathcal{M}$ generates the final answer $y$:
\begin{equation}
    y = \mathcal{M}(q, \text{Retrieve}(\mathcal{G}, q)). \label{eq1}
\end{equation}

The upper bound of generation quality heavily relies on $\mathcal{G}$'s structural quality. Traditionally, this is defined as a heterogeneous graph $\mathcal{G}_{trad} = (\mathcal{V}_E \cup \mathcal{V}_D, \mathcal{E})$, where $\mathcal{V}_E$ and $\mathcal{V}_D$ represent LLM-extracted entity nodes and document nodes, respectively. However, this exhibits two inherent flaws: (1) The massive entity set $\mathcal{V}_E$ incurs uncontrollable computational overhead; (2) Heuristic edge connections fail to rigorously balance relevance (intent alignment) and novelty (key document discovery).

To eliminate this overhead and precisely model relevance and novelty, we discard $\mathcal{V}_E$ and focus on a homogeneous document-only graph $\mathcal{G}$:

\begin{equation}
    \mathcal{G} = (\mathcal{V}, \mathcal{E}, \mathbf{W}), \label{eq2}
\end{equation}

\noindent where $\mathcal{V} = \{v_1, \dots, v_N\}$ is the document node set, $\mathcal{E} \subseteq \mathcal{V} \times \mathcal{V}$ denotes node edges, and $\mathbf{W} \in \mathbb{R}^{N \times N}$ is the adjacency weight matrix. Crucially, the edge weight $w_{ij}$ no longer relies on heuristic rules but is defined strictly as a functional mapping of the geometric relevance and novelty between $v_i$ and $v_j$ in a continuous vector space.

\section{Methodology}
\label{sec:methodology}

\subsection{Offline Geometric Graph Index}
The core of G$^3$RAG's graph construction lies in utilizing the geometric properties of document representations to establish topological connections. This fundamentally distinguishes it from frameworks like HippoRAG and LinearRAG, which incur high offline costs and struggle to balance relevance and novelty algorithmically.

Geometrically, the Relevance between two documents manifests as the directional consistency of their representations, measured by cosine similarity. For any nodes $v_i$ and $v_j$, we define their vector similarity as a scalar $x = \cos\theta \ (x \in [0, 1])$. In contrast, inherent Novelty implies minimal redundancy, meaning their representation directions diverge towards orthogonality. The information gain from this orthogonal component is characterized by $\sin\theta$. Using fundamental trigonometric identities, the novelty metric $\sin\theta$ is derived as follows:

\begin{equation}
    \sin\theta = \sqrt{1 - \cos^2\theta} = \sqrt{1 - x^2}. \label{eq3}
\end{equation}

After obtaining the cosine \textit{similarity-based} relevance metric $x$ and the \textit{orthogonal component-based} novelty metric $\sqrt{1-x^2}$, the subsequent core objective lies in complementary angular components for relevance-conditioned novelty computation. In complex multi-hop reasoning scenarios, an ideal document association link must anchor the current context to maintain thematic consistency (high relevance) while simultaneously exploring non-homogenous knowledge with significant information gain (high novelty). To achieve an endogenous balance between the two, we utilize the mutually restrictive nature of multiplicative operations to construct the edge weight $g_{ij}$ between graph nodes $v_i$ and $v_j$. Specifically, we define the edge weight as the product of relevance and novelty (i.e., $\cos\theta \cdot \sin\theta$), formally expressed as follows:

\begin{equation}
    g_{ij} = \cos\theta \cdot \sin\theta = x \cdot \sqrt{1 - x^2}. \label{eq4}
\end{equation}

Through a derivative analysis of the aforementioned equation, we observe that the edge weight function achieves its theoretical maximum when the cosine similarity is $x = \frac{\sqrt{2}}{2} \approx 0.707$ (the derivation process is shown in the appendix \ref{appendix:derivation}). Geometrically, this stationary point represents the optimal theoretical balance between relevance and novelty. Conversely, when the similarity is excessively low (e.g., $x < 0.1$), connections between nodes may exhibit semantic drift, making the system highly susceptible to topological noise that deviates from the core topic. Therefore, we strictly constrain the truncation interval of the similarity $x$ for valid edge connections to $[0.1, 0.7]$.

\subsection{Topological Degree Penalty}

After calculating the initial symmetric edge weights $g_{ij} = x\sqrt{1-x^2}$ based on geometric properties, we observe a topological risk: nodes with high similarity density (i.e., homogenized nodes) tend to impair node differentiation. To address this, we introduce a topological penalty mechanism in the final stage of graph construction. Specifically, when establishing a connection from a source node $v_i$ to a target node $v_j$, we first evaluate the global redundancy of the target node $v_j$, defined as its average similarity density estimate across the entire graph:

\begin{equation}
    \rho _j = \frac{1}{N} \sum_{k=1}^{N} \text{sim}(\mathbf{v}_j, \mathbf{v}_k),
\end{equation}
where $\mathbf{v}_j$ and $\mathbf{v}_k$ represent the continuous representation vectors of document nodes $v_j$ and $v_k$, respectively; $\text{sim}(\cdot, \cdot)$ denotes the cosine similarity function, and $N$ is the total number of nodes in the graph. Based on the obtained density estimate $\rho _j$, we apply an exponential decay function with a scaling hyperparameter $\beta$ to non-linearly modulate the initial edge weight $w'_{ij}$, thereby enforcing strict topological constraints:

\begin{equation}
    w_{ij} = w'_{ij} \cdot \exp(-\beta \cdot \rho _j), \label{eq5}
\end{equation}

Because $\rho_j$ measures average similarity to the corpus, larger values indicate greater similarity density and induce stronger penalties on edges entering $v_j$. This target-dependent modulation provides a hubness correction, conceptually related to local and global scaling, Mutual Proximity, and cross-domain similarity local scaling \cite{radovanovic2010hubs,schnitzer2012local,conneau2017word}. Applying it to the symmetric geometric weights combines density correction with relevance-conditioned novelty and yields generally asymmetric transition weights ($w_{ij} \neq w_{ji}$). This weighting encourages diffusion toward complementary evidence while reducing the influence of redundant hubs. The penalty-modulated weights form the adjacency matrix $\mathbf{W}$ of $\mathcal{G} = (\mathcal{V}, \mathcal{E}, \mathbf{W})$. Appendix \ref{sec:density_scaling} compares the global density estimate with a local alternative.

\subsection{Online Controlled Diffusion Inference}

Identifying high-quality seed nodes is critical for anchoring the multi-hop reasoning trajectory. While dense retrieval $\mathcal{R}_{\text{dense}}(q, k)$ can efficiently locate the "first-hop" evidence, it inherently suffers from the similarity trap, often retrieving false-positive neighbors that trigger cascading semantic drift during subsequent graph diffusion. To strictly enforce the purity of the source nodes, we employ a lightweight LLM-based reranker, denoted as $Filter_{\text{llm}}(\cdot)$, functioning as a hard gating mechanism:

\begin{equation}
    \mathcal{S} = Filter_{\text{llm}}(\mathcal{R}_{\text{dense}}(q, k), s), \label{eq6}
\end{equation}
where only the top-$s$ highest-confidence documents are retained to form the rigid seed set $\mathcal{S}$.Given the seed set $\mathcal{S}$, we construct an initial state vector $\mathbf{p}' \in \mathbb{R}^{|\mathcal{V}|}$, where $p'_i = 1/s$ if $v_i \in \mathcal{S}$, and $0$ otherwise. Unlike traditional random walk algorithms (e.g., PageRank) that require iterative message passing and often lead to topological hub absorption, G$^3$RAG performs an efficient Online Controlled Diffusion:

\begin{equation}
    \mathbf{p} = Rank(\mathbf{p}' \cdot \mathbf{T}, n) ,\label{eq7}
\end{equation}
where $\mathbf{T}$ represents the row-normalized transition matrix of the pre-built G$^3$RAG topology. Crucially, because the edge weights in $\mathbf{T}$ are strictly governed by our geometric information gain metric, this single-step diffusion $\mathbf{p}' \cdot \mathbf{T}$ is not a blind propagation. Instead, it acts as a directed geometric projection, naturally radiating activation energy toward target nodes that reside in orthogonal semantic subspaces (ensuring novelty) while bypassing topological echo chambers (via hub penalty). Finally, the top-$n$ documents derived from the probability distribution $\mathbf{p}$ are aggregated as the structured context for the reasoning model $\mathcal{M}$ to deduce the final answer $y$.

\section{Experiments}
\label{sec:experiment}
\subsection{Experimental Setup}
\textbf{Datasets.} We evaluated G$^3$RAG using 1,000 sampled cases from HippoRAG's curated versions of three QA datasets: MusiQue, 2WikiMultiHopQA, and HotpotQA \cite{gutierrez2024hipporag}. Among them, MuSiQue remains the most challenging due to more reasoning hops and concealed provenances. Notably, although these queries are typically associated with multiple intermediate supporting documents, the provenance document containing the final answer is strictly unique. This characteristic poses an exceptionally rigorous retrieval challenge: even if the system successfully retrieves an abundance of relevant intermediate clues, failing to reach this single terminal answer document will still result in the failure of the final question-answering process. We use the passage-level corpora released with HippoRAG2 \cite{gutierrez2025rag} without additional chunking or overlap. Each G$^3$RAG node contains a passage's title and body text. Mean document lengths are 82.7, 74.3, and 93.2 space-separated words for MuSiQue, 2WikiMultiHopQA, and HotpotQA, respectively. Corpus preprocessing is identical across embedding models, whose representation quality can affect G$^3$RAG's performance.

\textbf{Baselines.} We designed three categories of comparative baselines for a comprehensive evaluation: (1) LLM-Only: Utilizing Llama3.3-70B \cite{grattafiori2024llama3herdmodels}, Qwen3-32B (Thinking) \cite{yang2025qwen3technicalreport} to demonstrate the inherent difficulty of multi-hop queries and the necessity of external retrieval. (2) Naive RAG Methods: Encompassing dense retrieval approaches based on Nv-embed-v2 \cite{lee2024nv}, Qwen3-8B-embed \cite{zhang2025qwen3} to reveal the limitations of relying exclusively on surface-level semantic matching. (3) Structured Graph-based RAG Methods: Including SOTA baselines such as LightRAG \cite{guo2024lightrag}, RAPTOR \cite{sarthi2024raptor}, HippoRAG2 \cite{gutierrez2025rag}, and LinearRAG \cite{zhuang2025linearrag}. Crucially, to eliminate the impact of embedding preference on the advantages of our method, all RAG-related methods are evaluated and compared across two different embedding models: Nv-embed-v2 and Qwen3-8B-embed.

\textbf{Metrics.} We primarily report Exact Match (EM) and F1 scores for QA results, as they directly indicate the successful retrieval of the final answer document rather than mere intermediate clues. Regarding retrieval performance, we report Recall@5 and HitAnswer@5 (HitAns). HitAns@5 measures the success rate of retrieving the specific source document containing the final answer. In the main results, we report F1, Recall, and HitAns for comparison with other methods. In all other experiments, we report EM and F1, as these metrics directly reflect the final QA performance and the successful retrieval of the key answer document.

\textbf{Experimental Details.}  To ensure strict fairness, all RAG methods retrieve exactly 5 final documents and employ identical prompts for generation under each specified embedding model. During offline graph construction, LinearRAG uses its default BERT \cite{devlin2019bert} for entity extraction, while all other graph baselines use Llama-3.3-70B. During online inference, methods with LLM node filtering (HippoRAG2, LinearRAG, and G$^3$RAG) share aligned parameters: $k=10$ initial candidates, $s=5$ retained seed nodes (G$^3$RAG is $s=2$), and a final recall target of $n=5$. G$^3$RAG's specific topological penalty coefficient is set to $\beta=0.7$ for the MusiQue dataset and $\beta=0.1$ for the others. Detailed settings are provided in the appendix \ref{sec:appendix_A}.

\subsection{Main Results}
\label{sec:main results}

\textbf{Retrieval and Generation Results.} As shown in Table \ref{tab:results}, G$^3$RAG consistently outperforms baselines in average generation performance (F1 score) across both Nv-embed-v2 and Qwen3-8B-embed models. Under Nv-embed-v2, G$^3$RAG achieves an average F1 of 66.49, exceeding the strongest baseline, HippoRAG2 (64.49), by 2.00 points. Under Qwen3-8B-embed, the corresponding scores are 64.86 and 59.94, yielding a 4.92-point improvement based on the reported averages. On the most challenging MuSiQue dataset, G$^3$RAG exhibits even more substantial improvements, outperforming HippoRAG2 by 4.68 and 5.76 F1 points, respectively. Notably, these gains are achieved exclusively via G$^3$RAG's \textbf{zero-token} graph paradigm. Furthermore, LightRAG's underperformance compared to NaiveRAG corroborates that traditional graph methods are susceptible to the \textbf{"novelty trap"}, where off-topic noise degrades generation. Additionally, despite a relatively lower overall recall under Nv-embed-v2, G$^3$RAG maintains the highest key document hit rate (HitAns). It leads by an average of 0.43\% and 3.5\% across the two embedding models, and notably by 2.1\% and 5.5\% on the MuSiQue dataset. This indicates that HippoRAG2's document node constraints trap it in transitional bridge documents, missing distant novel keys. Conversely, G$^3$RAG anchors on bridge nodes while prioritizing novel document exploration. This advantage remains robust even under the less capable Qwen3-8B-embed model. This stable superiority fundamentally stems from G$^3$RAG's explicit focus on geometric gain: it rigorously evaluates novelty alongside similarity to encourage connections to novel documents. Grounded on this robust topology, G$^3$RAG leverages controlled transient diffusion to efficiently integrate optimal supporting evidence.

\textbf{Cost and Efficiency Results.} We evaluated the computational overhead of different RAG paradigms using Llama-3.3-70B and Nv-embed-v2 (batch size 4). As shown in Table \ref{tab:cost_consumption}, LightRAG consumed a staggering 86.80 M tokens to construct its massive graph, an unconstrained extraction incurring exorbitant costs and triggering the "Novelty Trap" during retrieval. Even HippoRAG2 required 14.47 M tokens. In stark contrast, G$^3$RAG achieves a true \textbf{"zero-token"} construction by computing edge weights solely via geometric gain, while consistently delivering optimal performance. Furthermore, although LinearRAG employs a relatively inexpensive BERT model for entity extraction, its index construction time remains excessively long and more GPU memory intensive. Additionally, limited by the BERT model, its graph construction robustness significantly underperforms compared to LLMs, with the second-best HippoRAG2 and G$^3$RAG exceeding its F1 performance on MuSiQue by 7.44 and 12.12 points, respectively. Regarding construction time and GPU memory, G$^3$RAG exhibits an absolute lightweight advantage, incurring zero memory overhead beyond the required embedding model. Conversely, other Graph RAG methods demand significant GPU resources to host powerful LLMs for graph extraction, hampering their real-world deployment. Benefiting from its minimalist architecture that requires no LLM during graph construction, G$^3$RAG facilitates ultra-lightweight knowledge establishment and ongoing maintenance. Given these advantages, alongside its highly efficient online inference characteristics, the framework demonstrates exceptional potential for practical deployment applications.

\begin{table*}[t]
    \centering
    \caption{Comparison of Performance. \textbf{Bold} indicates the best results, and \underline{underline} indicates the second-best. The metrics are divided into two groups based on Nv-embed-v2 and Qwen3-8B-embed for Graph-based RAG, evaluating the best and second-best performances within each group.}
    \label{tab:results}
    \resizebox{\textwidth}{!}{%
    \begin{tabular}{l *{12}{c}}
        \toprule
        \multirow{2}{*}{Methods} & \multicolumn{3}{c}{MuSiQue} & \multicolumn{3}{c}{2Wiki} & \multicolumn{3}{c}{Hotpotqa} & \multicolumn{3}{c}{Average} \\
        \cmidrule(lr){2-4} \cmidrule(lr){5-7} \cmidrule(lr){8-10} \cmidrule(lr){11-13}
        & F1 & Recall & HitAns & F1 & Recall & HitAns & F1 & Recall & HitAns & F1 & Recall & HitAns \\
        \midrule
        
        \rowcolor{gray!20} \multicolumn{13}{c}{\textbf{LLM-Only}} \\
        Qwen3-32B (Thinking) \cite{yang2025qwen3technicalreport} & 16.03 & - & - & 31.88 & - & - & 35.59 & - & - & 27.83 & - & - \\
        Llama3.3-70B \cite{grattafiori2024llama3herdmodels} & 26.73 & - & - & 47.51 & - & - & 48.98 & - & - & 41.07 & - & - \\

        \rowcolor{gray!20} \multicolumn{13}{c}{\textbf{Naive RAG}} \\
        Nv-embed-v2 \cite{lee2024nv} & 46.21 & 66.28 & 55.70 & 60.41 & 72.38 & 53.90 & 75.25 & 93.10 & 82.50 & 60.62 & 77.25 & 64.03 \\
        Qwen3-8B-embed \cite{zhang2025qwen3} & 44.35 & 61.08 & 53.90 & 62.37 & 72.02 & 54.70 & 71.77 & 87.90 & 77.40 & 59.50 & 73.67 & 62.00 \\

        \rowcolor{gray!20} \multicolumn{13}{c}{\textbf{Structured Graph-based RAG (Nv-embed-v2)}} \\
        RAPTOR \cite{sarthi2024raptor} & 28.90 & 59.23 & 51.20 & 52.10 & 69.35 & 53.30 & 69.50 & 70.65 & 79.70 & 50.17 & 66.41 & 61.40 \\
        Light RAG \cite{guo2024lightrag} & 23.09 & 28.16 & 27.83 & 24.56 & 43.18 & 33.60 & 39.27 & 35.49 & 45.68 & 28.97 & 35.61 & 35.70 \\
        HippoRAG2 \cite{gutierrez2024hipporag,gutierrez2025rag} & \underline{47.60} & \textbf{70.82} & \underline{63.10} & \textbf{70.57} & \textbf{89.65} & \textbf{72.20} & \underline{75.30} & \textbf{95.30} & \underline{85.90} & \underline{64.49} & \textbf{85.26} & \underline{73.73} \\
        LinearRAG \cite{zhuang2025linearrag} & 40.16 & 57.25 & 46.60 & 62.37 & 78.03 & 60.80 & 68.89 & 82.15 & 71.90 & 57.14 & 72.48 & 59.77 \\
        G$^3$RAG (Ours) & \textbf{52.28} & \underline{65.39} & \textbf{65.20} & \underline{69.20} & \underline{82.15} & \underline{70.60} & \textbf{78.00} & \underline{93.35} & \textbf{86.70} & \textbf{66.49} & \underline{80.30} & \textbf{74.17} \\

        \rowcolor{gray!20} \multicolumn{13}{c}{\textbf{Structured Graph-based RAG (Qwen3-8B-embed)}} \\
        RAPTOR \cite{sarthi2024raptor} & 38.90 & 59.19 & 52.90 & 56.24 & 71.17 & 54.60 & 67.80 & 69.35 & 76.60 & 54.31 & 66.57 & 61.37 \\
        Light RAG \cite{guo2024lightrag} & 18.20 & 24.67 & 26.41 & 20.97 & 53.17 & 45.20 & 39.27 & 35.49 & 45.68 & 26.15 & 37.78 & 39.10 \\
        HippoRAG2 \cite{gutierrez2024hipporag,gutierrez2025rag} & \underline{43.00} & \underline{60.94} & \underline{53.90} & 64.50 & 79.60 & 64.50 & \underline{72.33} & \textbf{91.15} & \underline{81.40} & \underline{59.94} & \textbf{77.23} & \underline{66.60} \\
        LinearRAG \cite{zhuang2025linearrag} & 39.71 & 55.82 & 45.60 & \underline{65.15} & \textbf{81.53} & \underline{66.30} & 68.40 & 81.30 & 73.10 & 57.75 & 72.88 & 61.67 \\
        G$^3$RAG (Ours) & \textbf{48.76} & \textbf{61.08} & \textbf{59.40} & \textbf{67.77} & \underline{79.88} & \textbf{67.60} & \textbf{78.04} & \underline{89.35} & \textbf{83.40} & \textbf{64.86} & \underline{76.77} & \textbf{70.13} \\

        \bottomrule
    \end{tabular}%
    }
\end{table*}

\begin{table}[t]
    \centering
    \caption{Cost and Consumption Comparison on MuSiQue Dataset with Llama-3.3-70B and Nv-embed-v2. $^\dagger$ indicates that the method uses a BERT-based entity extraction model.}
    \label{tab:cost_consumption}
    \footnotesize
    \resizebox{\columnwidth}{!}{%
    \begin{tabular}{l cccccc}
        \toprule
        Metric & NV-Embed-v2 & RAPTOR & LightRAG & LinearRAG & HippoRAG2 & G$^3$RAG \\
        \midrule
        
        Index Tokens (millions) & \textbf{0} & 1.90 & 86.80 & \underline{1.24 $^\dagger$} & 14.47 & \textbf{0} \\
        Index Time (min) & \textbf{$<$ 30} & $<$ 200 & $>$ 1440 & \underline{$<$ 100} & $<$ 300 & \textbf{$<$ 30} \\
        Index GPU Memory (GB) & \textbf{24} & 344 & 344 & \underline{30} & 344 & \textbf{24} \\
        QA time/Query (sec) & \textbf{0.30} & \underline{0.60} & 4.50 & 1.02 & 1.17 & 1.00 \\
        F1 & 46.21 & 28.90 & 23.09 & 40.16 & \underline{47.60} & \textbf{52.28} \\
        \bottomrule
    \end{tabular}%
    }
\end{table}

\begin{table*}[t]
    \centering
    \caption{Ablation Study Results}
    \label{tab:ablation}
    \small
    \resizebox{0.95\textwidth}{!}{%
    \begin{tabular}{l cccccccc}
        \toprule
        \multirow{2}{*}{Mode} & \multicolumn{2}{c}{MuSiQue} & \multicolumn{2}{c}{2Wiki} & \multicolumn{2}{c}{HotpotQA} & \multicolumn{2}{c}{Average} \\
        \cmidrule(lr){2-3} \cmidrule(lr){4-5} \cmidrule(lr){6-7} \cmidrule(lr){8-9}
        & EM & F1 & EM & F1 & EM & F1 & EM & F1 \\
        \midrule
        Naive          & 34.20 & 46.21 & 56.10 & 60.41 & 59.10 & 75.25 & 49.80 & 60.62 \\
        \quad w/ $Filter_{\text{llm}}$ & 36.70 & 50.27 & 58.10 & 63.42 & 61.40 & 76.47 & 52.07 & 63.39 \\
        \midrule
        G$^3$RAG             & \textbf{39.10} & \textbf{52.28} & \textbf{62.30} & \textbf{69.20} & \textbf{62.20} & \textbf{78.00} & \textbf{54.53} & \textbf{66.49} \\
         \quad w/o Penalty & 37.80$_{\downarrow 1.30}$ & 50.65$_{\downarrow 1.63}$ & 62.10$_{\downarrow 0.20}$ & 68.49$_{\downarrow 0.71}$ & 62.00$_{\downarrow 0.20}$ & 77.60$_{\downarrow 0.40}$ & 53.97$_{\downarrow 0.56}$ & 65.58$_{\downarrow 0.91}$ \\
         \quad w/ PPR      & 37.30$_{\downarrow 1.80}$ & 51.02$_{\downarrow 1.26}$ & 62.40$_{\uparrow 0.10}$ & 68.79$_{\downarrow 0.41}$ & 61.60$_{\downarrow 0.60}$ & 77.14$_{\downarrow 0.86}$ & 53.77$_{\downarrow 0.76}$ & 65.65$_{\downarrow 0.84}$ \\
         \quad w/o $Filter_{\text{llm}}$ & 33.70$_{\downarrow 5.40}$ & 45.22$_{\downarrow 7.06}$ & 57.30$_{\downarrow 5.00}$ & 63.03$_{\downarrow 6.17}$ & 59.10$_{\downarrow 3.10}$ & 73.33$_{\downarrow 4.67}$ & 50.03$_{\downarrow 4.50}$ & 60.53$_{\downarrow 5.96}$ \\

        \bottomrule
    \end{tabular}%
    }
\end{table*}

\subsection{Ablation Study}
The core innovation of our proposed G$^3$RAG architecture relies on two indispensable modules: the geometric gain-based Topological Degree Penalty mechanism and the Controlled Transient Diffusion. To deeply investigate the independent contributions of these two key components, we conducted detailed ablation studies, with the results presented in Table 3. First, when the topological degree penalty module is removed (w/o Penalty), the model exhibits significant performance degradation on the MuSiQue dataset, which features a higher reasoning span difficulty. When processing highly complex queries, the system needs to gather more novel evidence spanning local contexts. Without topological constraints, the retrieval chain may fall into a "similarity trap," tending to connect locally highly homogenized redundant documents, thereby missing the critical documents that truly contain high information gain. Second, replacing G$^3$RAG's transient diffusion with traditional Personalized PageRank (PPR) \cite{yang2024efficient} clearly declines QA performance across all datasets. G$^3$RAG's graph already precisely defines edge weights via the $\cos\theta \cdot \sin\theta$ geometric prior. PPR's multi-hop walks inevitably trigger over-smoothing, destroying this carefully constructed weight distribution. In contrast, our single-step controlled transient diffusion losslessly aggregates optimal evidence while significantly reducing online computational overhead.

Furthermore, even when the Naive baseline incorporates $Filter_{\text{llm}}$ for node filtering, it underperforms G$^3$RAG. This advantage stems from G$^3$RAG's ability to precisely anchor highly relevant documents using $Filter_{\text{llm}}$ and subsequently perform controlled diffusion along the topological graph to discover distant novel evidence, thereby completing the semantic chain—a vital step, as relying solely on LLM filtering cannot extend beyond the initial document scope to reach these distant key documents. Conversely, completely removing the $Filter_{\text{llm}}$ module from G$^3$RAG causes a severe performance drop. This further validates the necessity and rationality of our retrieval design: anchoring high-quality relevant starting points before driving controlled graph diffusion.

\subsection{Analysis}
\label{sec:analysis}

To further validate the robustness of the G$^3$RAG architecture and reveal the inner mechanisms of its core components, this section conducts an in-depth analysis covering three aspects: (1) Sensitivity analysis of the penalty coefficient $\beta$ to explore the boundary effects of topological constraints across different difficulty levels. (2) Joint configuration analysis to evaluate the combined impact of seed node set size ($s$) and target recall number ($n$) on the diffusion chain. (3) The scalability and potential of G$^3$RAG as a \textbf{single-step retrieval} paradigm to seamlessly migrate into an \textbf{iterative retrieval framework}. Additionally, qualitative case studies to demonstrate the retrieval behaviors of G$^3$RAG and other methods are provided in Appendix \ref{sec:appendix_C}.

\begin{figure*}[t]
    \centering
    \includegraphics[width=0.95\textwidth]{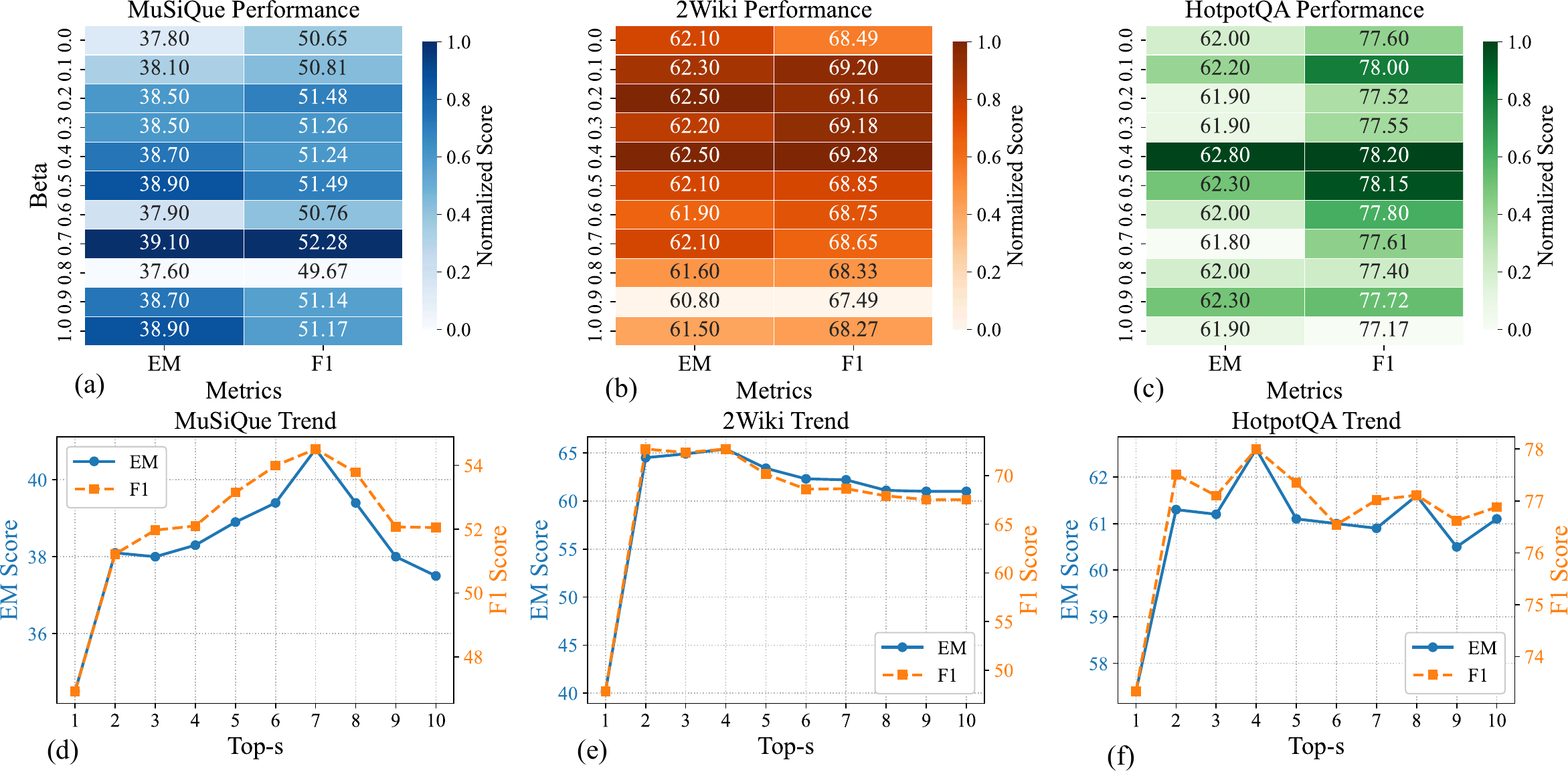}
    \caption{Parameter sensitivity. (a)-(c) present the sensitivity heatmaps of the $\beta$ parameter. (d)-(f) show the performance defects affected by the size of seed nodes top-$s$.}
    \label{fig:beta_sensitivity}
\end{figure*}

\textbf{Coefficient $\beta$ sensitivity analysis.} Figure \ref{fig:beta_sensitivity} (a-c) illustrate the framework's sensitivity to the topological penalty coefficient $\beta$. Consistently across datasets, applying a penalty ($\beta > 0$) strictly outperforms the unpenalized baseline ($\beta = 0.0$). This validates our motivation: without intervention, graph retrieval inherently traps reasoning in dense, uninformative hubs. Furthermore, optimal $\beta$ variations reflect differing dataset characteristics. For Wikipedia-centric corpora (HotpotQA, 2WikiMultiHopQA), an intermediate penalty ($\beta \approx 0.4$) is optimal, as many high-density nodes act as legitimate bridge entities crucial for logic continuation, over-penalizing them ($\beta \ge 0.8$) disrupts the reasoning chain. In contrast, the adversarially constructed MuSiQue dataset features densely entangled distractors mimicking central hubs, requiring a significantly stronger penalty ($\beta \approx 0.7$) to suppress deceptive nodes and force the retrieval of isolated, long-tail evidence.

\textbf{Joint configuration analysis of $s$ and $n$.} Figure \ref{fig:beta_sensitivity} (d-f) illustrates the impact of seed size $s$ under a fixed recall budget ($n=10$). The performance initially rises, as moderately increasing $s$ provides robust semantic anchors. However, it degrades once $s$ crosses a threshold (e.g., $s \ge 4$ for 2Wiki and HotpotQA). This is primarily due to the "Budget Crowding-out Effect": excessive seeds deplete the limited retrieval window, compressing the topological space for outward diffusion and preventing the recall of multi-hop novel evidence. Notably, the MuSiQue dataset, with its high combinatorial interference, requires more seeds to anchor reasoning, delaying the peak to $s=7$. Beyond this ($s>7$), the severely restricted remaining diffusion quota fails to accommodate new evidence, causing a performance drop.

\begin{table}[t]
    \centering
    \caption{Performance Comparison of Iterative Methods. G$^3$RAG is embedded into the iterative framework as a retrieval method.}
    \label{tab:iterative_results}
    \resizebox{0.98\columnwidth}{!}{%
    \begin{tabular}{l cccccccc}
        \toprule
        \multirow{2}{*}{Mode} & \multicolumn{2}{c}{MuSiQue} & \multicolumn{2}{c}{2Wiki} & \multicolumn{2}{c}{Hotpotqa} & \multicolumn{2}{c}{Average} \\
        \cmidrule(lr){2-3} \cmidrule(lr){4-5} \cmidrule(lr){6-7} \cmidrule(lr){8-9}
        & EM & F1 & EM & F1 & EM & F1 & EM & F1 \\
        \midrule
        
        IterDRAG \cite{yue2024inference}   & 34.90 & 48.44 & 56.20 & 63.83 & 59.90 & 75.34 & 50.33 & 62.54 \\
        \quad w/ G$^3$RAG      & 39.20$_{\uparrow 4.30}$ & 52.01$_{\uparrow 3.57}$ & 60.20$_{\uparrow 4.00}$ & 69.04$_{\uparrow 5.21}$ & 62.20$_{\uparrow 2.30}$ & 77.74$_{\uparrow 2.40}$ & 53.87$_{\uparrow 3.54}$ & 66.26$_{\uparrow 3.72}$ \\
        \addlinespace 
        
        IterRetgen \cite{shao2023enhancing} & 39.20 & 51.71 & 59.10 & 64.12 & 61.70 & 76.91 & 53.33 & 64.25 \\
        \quad w/ G$^3$RAG      & 41.90$_{\uparrow 2.70}$ & 54.54$_{\uparrow 2.83}$ & 64.80$_{\uparrow 5.70}$ & 72.63$_{\uparrow 8.51}$ & 62.60$_{\uparrow 0.90}$ & 78.09$_{\uparrow 1.18}$ & 56.43$_{\uparrow 3.10}$ & 68.42$_{\uparrow 4.17}$ \\
        \addlinespace 

        \bottomrule
    \end{tabular}%
    }
\end{table}

\textbf{Iterative Integration.} G$^3$RAG seamlessly integrates into iterative workflows. In Table \ref{tab:iterative_results}, incorporating G$^3$RAG significantly enhances IterDRAG \cite{yue2024inference} and IterRetgen \cite{shao2023enhancing}. This improvement occurs because iterative approaches rely on the precise retrieval of key documents, a task where G$^3$RAG's geometric gain graph excels. Furthermore, although iterative paradigms decompose queries, some sub-questions may involve multi-hop difficulty and still require novel evidence for support. This amply illustrates that G$^3$RAG, driven by its distinctive knowledge construction methodology and retrieval strategy, exhibits remarkable scalability, thereby demonstrating enormous potential for real-world deployment within mainstream RAG and Agent systems.

\textbf{Graph search strategies.} To examine the interaction between graph construction and online retrieval, we keep the constructed graph fixed and replace controlled transient diffusion with Personalized PageRank (PPR), greedy walk, or beam search. As shown in Table \ref{tab:graph_search_strategies}, G$^3$RAG achieves the highest average EM and F1 scores of 54.53 and 66.49, respectively. Its average F1 exceeds those of PPR, greedy walk, and beam search by 0.84, 1.70, and 1.62 points, respectively, indicating that the choice of search strategy affects how effectively the graph supports evidence retrieval. The observed differences are consistent with the risks associated with these search mechanisms. Greedy walk commits to the highest-scoring neighbor at each step, which can propagate local selection errors and introduce false positives. Beam search explores a broader candidate space, potentially admitting irrelevant documents, while PPR's repeated propagation can spread activation to noisy nodes beyond the seed neighborhoods. G$^3$RAG instead performs a single diffusion step from the selected seed nodes, directly using the geometric gain encoded in the offline edge weights to retrieve complementary evidence. These results support the joint design of geometric graph construction and controlled transient diffusion within a framework that incurs no LLM token overhead during graph construction. The alternative search strategies nevertheless retain average F1 scores of 64.79--65.65, exceeding both Naive (60.62) and HippoRAG2 (64.49). This advantage suggests that the constructed graph remains useful across different search strategies, supporting its robustness and adaptability. The advantage over HippoRAG2 is specific to average F1, as the alternative strategies yield lower average EM scores.

\begin{table}[t]
    \centering
    \caption{Comparison of graph search strategies using Nv-embed-v2 as the embedding model. PPR, Greedy Walk, and Beam Search use the same graph as G$^3$RAG.}
    \label{tab:graph_search_strategies}
    \small
    \resizebox{0.88\columnwidth}{!}{%
    \begin{tabular}{l cccccccc}
        \toprule
        \multirow{2}{*}{Mode} & \multicolumn{2}{c}{MuSiQue} & \multicolumn{2}{c}{2Wiki} & \multicolumn{2}{c}{HotpotQA} & \multicolumn{2}{c}{Average} \\
        \cmidrule(lr){2-3} \cmidrule(lr){4-5} \cmidrule(lr){6-7} \cmidrule(lr){8-9}
        & EM & F1 & EM & F1 & EM & F1 & EM & F1 \\
        \midrule
        Naive       & 34.20 & 46.21 & 56.10 & 60.41 & 59.10 & 75.25 & 49.80 & 60.62 \\
        HippoRAG2 \cite{gutierrez2024hipporag,gutierrez2025rag}  & 36.70 & 47.60 & \textbf{63.80} & \textbf{70.57} & \textbf{62.40} & 75.30 & 54.30 & 64.49 \\
        \midrule
        \quad PPR \cite{page1998pagerank}        & 37.30 & 51.02 & 62.40 & 68.79 & 61.60 & 77.14 & 53.77 & 65.65 \\
        \quad Greedy Walk \cite{kleinberg2000small} & 37.90 & 51.16 & 60.10 & 65.86 & 62.00 & 77.34 & 53.33 & 64.79 \\
        \quad Beam Search \cite{lowerre1976harpy} & 37.80 & 50.80 & 60.70 & 66.29 & 61.80 & 77.53 & 53.43 & 64.87 \\
        G$^3$RAG    & \textbf{39.10} & \textbf{52.28} & 62.30 & 69.20 & 62.20 & \textbf{78.00} & \textbf{54.53} & \textbf{66.49} \\
        \bottomrule
    \end{tabular}%
    }
\end{table}

\section{Conclusions}
\label{sec:conlude}

This paper introduces GeoGainGraph (G$^3$RAG), a zero-token graph construction paradigm designed to address the computational and performance bottlenecks of multi-hop RAG systems. By modeling the relevance-novelty trade-off via directional consistency ($\cos\theta$) and orthogonality ($\sin\theta$), G$^3$RAG constructs a robust topology without expensive LLMs, while penalizing dense nodes to uncover long-tail novel documents. Based on this geometric graph, the proposed controlled transient diffusion mechanism precisely anchors seed nodes to dynamically retrieve the optimal combination of multi-hop evidence. Extensive experimental results demonstrate that, despite its zero-token construction approach, G$^3$RAG consistently achieves state-of-the-art performance. Although G$^3$RAG demonstrates significant advantages as outlined above, establishing a universal numerical threshold for the optimal geometric gain remains a considerable challenge. However, this challenge reinforces a core insight of our framework: the practical significance of geometric gain lies in its directionality rather than rigid numerical constraints. By continuously steering the retrieval trajectory away from both semantic redundancy and topic drift, our method maintains high robustness across diverse underlying models. Building on this directional philosophy, future work will generalize our geometric gain mechanism to multi-modal reasoning scenarios, leveraging orthogonal information gain to align complementary knowledge across distinct modal spaces.

\section*{Acknowledgements}

This work is supported by the Key-Area Research and Development Program of Guangdong Province (Grant No.~2024B0101040004) and the 2025 Guangzhou Industry Collaborative Innovation Center Construction Special Funding (Grant No.~SL2024B04J00047). This work is also partially supported by the High Performance Computing Platform of South China University of Technology.

\clearpage
\bibliographystyle{plain}
\bibliography{custom}

@article{lewis2020retrieval,
  title={Retrieval-augmented generation for knowledge-intensive nlp tasks},
  author={Lewis, Patrick and Perez, Ethan and Piktus, Aleksandra and Petroni, Fabio and Karpukhin, Vladimir and Goyal, Naman and K{\"u}ttler, Heinrich and Lewis, Mike and Yih, Wen-tau and Rockt{\"a}schel, Tim and others},
  journal={Advances in neural information processing systems},
  volume={33},
  pages={9459--9474},
  year={2020}
}

@inproceedings{li-etal-2025-unilr,
    title = "{U}ni{LR}: Unleashing the Power of {LLM}s on Multiple Legal Tasks with a Unified Legal Retriever",
    author = "Li, Ang  and
      Wu, Yiquan  and
      Liu, Yifei  and
      Cai, Ming  and
      Qing, Lizhi  and
      Wang, Shihang  and
      Kang, Yangyang  and
      Liu, Chengyuan  and
      Wu, Fei  and
      Kuang, Kun",
    editor = "Che, Wanxiang  and
      Nabende, Joyce  and
      Shutova, Ekaterina  and
      Pilehvar, Mohammad Taher",
    booktitle = "Proceedings of the 63rd Annual Meeting of the Association for Computational Linguistics (Volume 1: Long Papers)",
    month = jul,
    year = "2025",
    address = "Vienna, Austria",
    publisher = "Association for Computational Linguistics",
    url = "https://aclanthology.org/2025.acl-long.584/",
    doi = "10.18653/v1/2025.acl-long.584",
    pages = "11953--11967",
    ISBN = "979-8-89176-251-0"
}

@inproceedings{tan-etal-2025-prospect,
    title = "In Prospect and Retrospect: Reflective Memory Management for Long-term Personalized Dialogue Agents",
    author = "Tan, Zhen  and
      Yan, Jun  and
      Hsu, I-Hung  and
      Han, Rujun  and
      Wang, Zifeng  and
      Le, Long  and
      Song, Yiwen  and
      Chen, Yanfei  and
      Palangi, Hamid  and
      Lee, George  and
      Iyer, Anand Rajan  and
      Chen, Tianlong  and
      Liu, Huan  and
      Lee, Chen-Yu  and
      Pfister, Tomas",
    editor = "Che, Wanxiang  and
      Nabende, Joyce  and
      Shutova, Ekaterina  and
      Pilehvar, Mohammad Taher",
    booktitle = "Proceedings of the 63rd Annual Meeting of the Association for Computational Linguistics (Volume 1: Long Papers)",
    month = jul,
    year = "2025",
    address = "Vienna, Austria",
    publisher = "Association for Computational Linguistics",
    url = "https://aclanthology.org/2025.acl-long.413/",
    doi = "10.18653/v1/2025.acl-long.413",
    pages = "8416--8439",
    ISBN = "979-8-89176-251-0"
}

@misc{peng2024graphretrievalaugmentedgenerationsurvey,
      title={Graph Retrieval-Augmented Generation: A Survey}, 
      author={Boci Peng and Yun Zhu and Yongchao Liu and Xiaohe Bo and Haizhou Shi and Chuntao Hong and Yan Zhang and Siliang Tang},
      year={2024},
      eprint={2408.08921},
      archivePrefix={arXiv},
      primaryClass={cs.AI},
      url={https://arxiv.org/abs/2408.08921}, 
}

@article{zhang2025survey,
  title={A survey of graph retrieval-augmented generation for customized large language models},
  author={Zhang, Qinggang and Chen, Shengyuan and Bei, Yuanchen and Yuan, Zheng and Zhou, Huachi and Hong, Zijin and Dong, Junnan and Chen, Hao and Chang, Yi and Huang, Xiao},
  journal={arXiv preprint arXiv:2501.13958},
  year={2025}
}

@article{zhuang2024efficientrag,
  title={Efficientrag: Efficient retriever for multi-hop question answering},
  author={Zhuang, Ziyuan and Zhang, Zhiyang and Cheng, Sitao and Yang, Fangkai and Liu, Jia and Huang, Shujian and Lin, Qingwei and Rajmohan, Saravan and Zhang, Dongmei and Zhang, Qi},
  journal={arXiv preprint arXiv:2408.04259},
  year={2024}
}

@article{lee2024hybgrag,
  title={HybGrag: Hybrid retrieval-augmented generation on textual and relational knowledge bases},
  author={Lee, Meng-Chieh and Zhu, Qi and Mavromatis, Costas and Han, Zhen and Adeshina, Soji and Ioannidis, Vassilis N and Rangwala, Huzefa and Faloutsos, Christos},
  journal={arXiv preprint arXiv:2412.16311},
  year={2024}
}

@article{edge2024local,
  title={From local to global: A graph rag approach to query-focused summarization},
  author={Edge, Darren and Trinh, Ha and Cheng, Newman and Bradley, Joshua and Chao, Alex and Mody, Apurva and Truitt, Steven and Metropolitansky, Dasha and Ness, Robert Osazuwa and Larson, Jonathan},
  journal={arXiv preprint arXiv:2404.16130},
  year={2024}
}

@article{guo2024lightrag,
  title={Lightrag: Simple and fast retrieval-augmented generation},
  author={Guo, Zirui and Xia, Lianghao and Yu, Yanhua and Ao, Tu and Huang, Chao},
  journal={arXiv preprint arXiv:2410.05779},
  year={2024}
}

@article{gutierrez2025rag,
  title={From rag to memory: Non-parametric continual learning for large language models},
  author={Guti{\'e}rrez, Bernal Jim{\'e}nez and Shu, Yiheng and Qi, Weijian and Zhou, Sizhe and Su, Yu},
  journal={Forty-Second International Conference on Machine Learning},
  year={2025}
}

@article{zhao2024meta,
  title={Meta-Chunking: Learning Text Segmentation and Semantic Completion via Logical Perception},
  author={Zhao, Jihao and Ji, Zhiyuan and Feng, Yuchen and Qi, Pengnian and Niu, Simin and Tang, Bo and Xiong, Feiyu and Li, Zhiyu},
  journal={arXiv preprint arXiv:2410.12788},
  year={2024}
}

@article{zhuang2025linearrag,
  title={Linearrag: Linear graph retrieval augmented generation on large-scale corpora},
  author={Zhuang, Luyao and Chen, Shengyuan and Xiao, Yilin and Zhou, Huachi and Zhang, Yujing and Chen, Hao and Zhang, Qinggang and Huang, Xiao},
  journal={The Fourteenth International Conference on Learning Representations},
  year={2026}
}

@article{el2024cellular,
  title={A cellular basis for mapping behavioural structure},
  author={El-Gaby, Mohamady and Harris, Adam Loyd and Whittington, James CR and Dorrell, William and Bhomick, Arya and Walton, Mark E and Akam, Thomas and Behrens, Timothy EJ},
  journal={Nature},
  volume={636},
  number={8043},
  pages={671--680},
  year={2024},
  publisher={Nature Publishing Group UK London}
}

@inproceedings{zhu2025mitigating,
  title={Mitigating lost-in-retrieval problems in retrieval augmented multi-hop question answering},
  author={Zhu, Rongzhi and Liu, Xiangyu and Sun, Zequn and Wang, Yiwei and Hu, Wei},
  booktitle={Proceedings of the 63rd Annual Meeting of the Association for Computational Linguistics (Volume 1: Long Papers)},
  pages={22362--22375},
  year={2025}
}

@inproceedings{shen2025gear,
  title={Gear: Graph-enhanced agent for retrieval-augmented generation},
  author={Shen, Zhili and Diao, Chenxin and Vougiouklis, Pavlos and Merita, Pascual and Piramanayagam, Shriram and Chen, Enting and Graux, Damien and Melo, Andre and Lai, Ruofei and Jiang, Zeren and others},
  booktitle={Findings of the Association for Computational Linguistics: ACL 2025},
  pages={12049--12072},
  year={2025}
}

@inproceedings{glass2022re2g,
  title={Re2G: Retrieve, rerank, generate},
  author={Glass, Michael and Rossiello, Gaetano and Chowdhury, Md Faisal Mahbub and Naik, Ankita and Cai, Pengshan and Gliozzo, Alfio},
  booktitle={Proceedings of the 2022 Conference of the North American Chapter of the Association for Computational Linguistics: Human Language Technologies},
  pages={2701--2715},
  year={2022}
}

@article{wang2024corag,
  title={Corag: A cost-constrained retrieval optimization system for retrieval-augmented generation},
  author={Wang, Ziting and Yuan, Haitao and Dong, Wei and Cong, Gao and Li, Feifei},
  journal={arXiv preprint arXiv:2411.00744},
  year={2024}
}

@inproceedings{guo2025empowering,
  title={Empowering graphrag with knowledge filtering and integration},
  author={Guo, Kai and Shomer, Harry and Zeng, Shenglai and Han, Haoyu and Wang, Yu and Tang, Jiliang},
  booktitle={Proceedings of the 2025 Conference on Empirical Methods in Natural Language Processing},
  pages={25450--25464},
  year={2025}
}

@inproceedings{li2025t,
  title={T-grag: A dynamic graphrag framework for resolving temporal conflicts and redundancy in knowledge retrieval},
  author={Li, Dong and Niu, Yichen and Ai, Ying and Zou, Xiang and Qi, Biqing and Liu, Jianxing},
  booktitle={Proceedings of the 33rd ACM International Conference on Multimedia},
  pages={11880--11889},
  year={2025}
}

@article{han2025rag,
  title={Rag vs. graphrag: A systematic evaluation and key insights},
  author={Han, Haoyu and Ma, Li and Shomer, Harry and Wang, Yu and Lei, Yongjia and Guo, Kai and Hua, Zhigang and Long, Bo and Liu, Hui and Aggarwal, Charu C and others},
  journal={arXiv preprint arXiv:2502.11371},
  year={2025}
}

@inproceedings{you2025ms,
  title={MS-RAG: Simple and Effective Multi-Semantic Retrieval-Augmented Generation},
  author={You, Xiaozhou and Luo, Yahui and Gu, Lihong},
  booktitle={Proceedings of the 2025 Conference on Empirical Methods in Natural Language Processing},
  pages={22620--22636},
  year={2025}
}

@article{zhang2025qwen3,
  title={Qwen3 embedding: Advancing text embedding and reranking through foundation models},
  author={Zhang, Yanzhao and Li, Mingxin and Long, Dingkun and Zhang, Xin and Lin, Huan and Yang, Baosong and Xie, Pengjun and Yang, An and Liu, Dayiheng and Lin, Junyang and others},
  journal={arXiv preprint arXiv:2506.05176},
  year={2025}
}

@article{lee2024nv,
  title={Nv-embed: Improved techniques for training llms as generalist embedding models},
  author={Lee, Chankyu and Roy, Rajarshi and Xu, Mengyao and Raiman, Jonathan and Shoeybi, Mohammad and Catanzaro, Bryan and Ping, Wei},
  journal={arXiv preprint arXiv:2405.17428},
  year={2024}
}

@misc{grattafiori2024llama3herdmodels,
      title={The Llama 3 Herd of Models}, 
      author={Aaron Grattafiori and Abhimanyu Dubey and Abhinav Jauhri and Abhinav Pandey and Abhishek Kadian and others},
      year={2024},
      eprint={2407.21783},
      archivePrefix={arXiv},
      primaryClass={cs.AI},
      url={https://arxiv.org/abs/2407.21783}, 
}

@article{gutierrez2024hipporag,
  title={Hipporag: Neurobiologically inspired long-term memory for large language models},
  author={Guti{\'e}rrez, Bernal J and Shu, Yiheng and Gu, Yu and Yasunaga, Michihiro and Su, Yu},
  journal={Advances in neural information processing systems},
  volume={37},
  pages={59532--59569},
  year={2024}
}

@inproceedings{sarthi2024raptor,
  title={Raptor: Recursive abstractive processing for tree-organized retrieval},
  author={Sarthi, Parth and Abdullah, Salman and Tuli, Aditi and Khanna, Shubh and Goldie, Anna and Manning, Christopher D},
  booktitle={The Twelfth International Conference on Learning Representations},
  year={2024}
}

@misc{yang2025qwen3technicalreport,
      title={Qwen3 Technical Report}, 
      author={An Yang and Anfeng Li and Baosong Yang and Beichen Zhang and Binyuan Hui and Bo Zheng and Bowen Yu and Chang Gao and Chengen Huang and Chenxu Lv and Chujie Zheng and Dayiheng Liu and Fan Zhou and Fei Huang and Feng Hu and Hao Ge and Haoran Wei and Huan Lin and Jialong Tang and Jian Yang and Jianhong Tu and Jianwei Zhang and Jianxin Yang and Jiaxi Yang and Jing Zhou and Jingren Zhou and Junyang Lin and Kai Dang and Keqin Bao and Kexin Yang and Le Yu and Lianghao Deng and Mei Li and Mingfeng Xue and Mingze Li and Pei Zhang and Peng Wang and Qin Zhu and Rui Men and Ruize Gao and Shixuan Liu and Shuang Luo and Tianhao Li and Tianyi Tang and Wenbiao Yin and Xingzhang Ren and Xinyu Wang and Xinyu Zhang and Xuancheng Ren and Yang Fan and Yang Su and Yichang Zhang and Yinger Zhang and Yu Wan and Yuqiong Liu and Zekun Wang and Zeyu Cui and Zhenru Zhang and Zhipeng Zhou and Zihan Qiu},
      year={2025},
      eprint={2505.09388},
      archivePrefix={arXiv},
      primaryClass={cs.CL},
      url={https://arxiv.org/abs/2505.09388}, 
}

@inproceedings{devlin2019bert,
  title={Bert: Pre-training of deep bidirectional transformers for language understanding},
  author={Devlin, Jacob and Chang, Ming-Wei and Lee, Kenton and Toutanova, Kristina},
  booktitle={Proceedings of the 2019 conference of the North American chapter of the association for computational linguistics: human language technologies, volume 1 (long and short papers)},
  pages={4171--4186},
  year={2019}
}

@article{yang2024efficient,
  title={Efficient algorithms for personalized pagerank computation: A survey},
  author={Yang, Mingji and Wang, Hanzhi and Wei, Zhewei and Wang, Sibo and Wen, Ji-Rong},
  journal={IEEE Transactions on Knowledge and Data Engineering},
  volume={36},
  number={9},
  pages={4582--4602},
  year={2024},
  publisher={IEEE}
}

@article{yue2024inference,
  title={Inference scaling for long-context retrieval augmented generation},
  author={Yue, Zhenrui and Zhuang, Honglei and Bai, Aijun and Hui, Kai and Jagerman, Rolf and Zeng, Hansi and Qin, Zhen and Wang, Dong and Wang, Xuanhui and Bendersky, Michael},
  journal={The Thirteenth International Conference on Learning Representations},
  year={2025}
}

@inproceedings{shao2023enhancing,
  title={Enhancing retrieval-augmented large language models with iterative retrieval-generation synergy},
  author={Shao, Zhihong and Gong, Yeyun and Shen, Yelong and Huang, Minlie and Duan, Nan and Chen, Weizhu},
  booktitle={Findings of the Association for Computational Linguistics: EMNLP 2023},
  pages={9248--9274},
  year={2023}
}

@inproceedings{page1998pagerank,
  title={The pagerank citation ranking: Bring order to the web},
  author={Page, Lawrence and Brin, Sergey and Motwani, Rajeev and Winograd, Terry},
  booktitle={Proc. of the 7th International World Wide Web Conf},
  year={1998}
}

@inproceedings{kleinberg2000small,
  title={The small-world phenomenon: An algorithmic perspective},
  author={Kleinberg, Jon},
  booktitle={Proceedings of the thirty-second annual ACM symposium on Theory of computing},
  pages={163--170},
  year={2000}
}

@book{lowerre1976harpy,
  title={The harpy speech recognition system.},
  author={Lowerre, Bruce T},
  year={1976},
  publisher={Carnegie Mellon University}
}

@inproceedings{carbonell1998use,
  title={The use of MMR, diversity-based reranking for reordering documents and producing summaries.},
  author={Carbonell, Jaime G and Goldstein, Jade},
  booktitle={SIGIR},
  volume={98},
  number={10.1145},
  pages={290941--291025},
  year={1998}
}

@article{kulesza2012determinantal,
  title={Determinantal point processes for machine learning},
  author={Kulesza, Alex and Taskar, Ben},
  journal={Foundations and Trends{\textregistered} in Machine Learning},
  volume={5},
  number={2-3},
  pages={123--286},
  year={2012},
  publisher={Emerald Publishing Limited}
}

@article{radovanovic2010hubs,
  title={Hubs in space: Popular nearest neighbors in high-dimensional data},
  author={Radovanovic, Milos and Nanopoulos, Alexandros and Ivanovic, Mirjana},
  journal={Journal of machine learning research},
  volume={11},
  number={sept},
  pages={2487--2531},
  year={2010}
}

@article{schnitzer2012local,
  title={Local and global scaling reduce hubs in space.},
  author={Schnitzer, Dominik and Flexer, Arthur and Schedl, Markus and Widmer, Gerhard},
  journal={Journal of Machine Learning Research},
  volume={13},
  number={10},
  year={2012}
}

@article{conneau2017word,
  title={Word translation without parallel data},
  author={Conneau, Alexis and Lample, Guillaume and Ranzato, Marc'Aurelio and Denoyer, Ludovic and J{\'e}gou, Herv{\'e}},
  journal={arXiv preprint arXiv:1710.04087},
  year={2017}
}

\clearpage
\appendix

\section{Experimental Details}
\label{sec:appendix_A}

\paragraph{Hyperparameter Settings.} In addition to the core hyperparameters $k$, $s$, $n$, and $\beta$ detailed in the main text, we strictly followed the official recommended default configurations for all other structured RAG baselines. The specific key parameter settings are as follows: For HippoRAG2, the initial retrieval pool contains 200 documents. Its internal linking and filtering procedure selects $k=10$ candidates from this pool, and the top five documents are used for evaluation. The document node weight during the Personalized PageRank (PPR) walk is 0.05, the damping factor is 0.5, and the synonym edge threshold during graph construction is 0.8. For LinearRAG, the document node ratio during the PPR process is 1.5 with a weight of 0.05, and entities with a similarity below the 0.5 threshold during iteration are pruned. LightRAG limits the maximum summary tokens to 500 during graph construction and utilizes the "Local" mode for retrieval. Furthermore, RAPTOR's beam pruning threshold during tree construction is set to 0.5, the number of tree connections is 5, the maximum summary length is 100 tokens, and the maximum tree depth is set to 5 levels.

\paragraph{Prompt Settings.} For all baseline methods involving structured graph construction, the prompts are kept strictly consistent with their original papers. Furthermore, to ensure a uniform final output format and a fair evaluation, the prompt templates for final reasoning and answer generation are uniformly standardized to those provided by HippoRAG2 across all methods.

\paragraph{Computational Resources and Model Deployment.} All experiments were conducted in a standardized hardware environment. Specifically, for the generation and reasoning phase, we deployed the Llama3.3-70B large language model across 4 NVIDIA A800 GPUs. For the embedding phase, we utilized a single NVIDIA RTX 4090 GPU to deploy either the NV-Embed-v2 or Qwen3-8B-embed models. Regarding inference acceleration, with the exception of NV-Embed-v2, all models (including Llama3.3-70B and Qwen3-8B-embed) were deployed using the vLLM framework to maximize system throughput. During the offline graph construction for all baseline methods, because NV-Embed-v2 lacks vLLM acceleration support, its embedding batch size was strictly set to 4 due to memory allocation. In contrast, benefiting from underlying vLLM optimizations, Qwen3-8B-embed processes inputs efficiently with a batch size of 64.

\section{Algorithm Pseudocode}

In this section, we present the detailed pseudocode for the proposed GeoGainGraph (G$^3$RAG) framework. Algorithm \ref{alg:offline} outlines the offline geometric graph construction process, detailing the calculation of edge weights based on geometric information gain and the application of the topological degree penalty. Algorithm \ref{alg:online} delineates the online controlled diffusion inference, demonstrating the complete pipeline from anchoring reliable seed nodes to target evidence retrieval and final answer generation.
\begin{algorithm*}[t]
\caption{Offline Geometric Graph Construction}
\label{alg:offline}
\begin{algorithmic}[1]
\REQUIRE Document set $\mathcal{V} = \{v_1, v_2, \dots, v_N\}$, penalty hyperparameter $\beta$, bounds $[a, b] = [0.1, 0.7]$.
\ENSURE Adjacency weight matrix $\mathbf{W}$.
\STATE Initialize $\mathbf{G} \in \mathbb{R}^{N \times N}$ and $\mathbf{W} \in \mathbb{R}^{N \times N}$.
\FOR{each pair of documents $(v_i, v_j) \in \mathcal{V} \times \mathcal{V}$}
    \STATE $x \leftarrow \cos\theta = \text{sim}(\mathbf{v}_i, \mathbf{v}_j)$
    \IF{$x \in [a, b]$}
        \STATE $g_{ij} \leftarrow x \cdot \sqrt{1 - x^2}$
    \ELSE
        \STATE $g_{ij} \leftarrow 0$
    \ENDIF
    \STATE $\mathbf{G}_{i,j} \leftarrow g_{ij}$
\ENDFOR
\FOR{each target document $v_j \in \mathcal{V}$}
    \STATE $\rho_j \leftarrow \frac{1}{N} \sum_{k=1}^{N} \text{sim}(\mathbf{v}_j, \mathbf{v}_k)$
\ENDFOR
\FOR{each pair of documents $(v_i, v_j) \in \mathcal{V} \times \mathcal{V}$}
    \STATE $w_{ij} \leftarrow \mathbf{G}_{i,j} \cdot \exp(-\beta \cdot \rho_j)$
    \STATE $\mathbf{W}_{i,j} \leftarrow w_{ij}$
\ENDFOR
\RETURN $\mathbf{W}$
\end{algorithmic}
\end{algorithm*}
\begin{algorithm*}[t]
\caption{Online Controlled Diffusion Inference}
\label{alg:online}
\begin{algorithmic}[1]
\REQUIRE User query $q$, Graph $\mathcal{G}=(\mathcal{V}, \mathcal{E}, \mathbf{W})$, initial candidates $k$, seed size $s$, target recall $n$ ($1\leq s\leq\min(k,n)$), language model $\mathcal{M}$.
\ENSURE Generated answer $y$.
\STATE $\mathcal{R}_{\text{dense}} \leftarrow \text{Retrieve}(q, k)$ \COMMENT{Dense retrieval for initial candidates}
\STATE $\mathcal{S} \leftarrow Filter_{\text{llm}}(\mathcal{R}_{\text{dense}}, s)$ \COMMENT{LLM filtering for reliable seeds}
\STATE Initialize state vector $\mathbf{p}' \in \mathbb{R}^N$ with $0$
\FOR{$v_i \in \mathcal{S}$}
    \STATE $\mathbf{p}'_i \leftarrow 1 / s$
\ENDFOR
\STATE Compute transition matrix $\mathbf{T}$ from $\mathbf{W}$ (e.g., row normalization)
\STATE $\mathbf{p} \leftarrow \mathbf{p}' \cdot \mathbf{T}$ \COMMENT{Transient diffusion}
\begingroup
\STATE $\mathcal{D}_{ret} \leftarrow \mathcal{S}$ \COMMENT{Retain all seed documents}
\IF{$s<n$}
    \STATE $\mathcal{A} \leftarrow \operatorname{Top}_{n-s}\{v_i \in \mathcal{V}\setminus\mathcal{S}\text{ ranked by }p_i\}$
    \STATE $\mathcal{D}_{ret} \leftarrow \mathcal{S}\cup\mathcal{A}$ \COMMENT{Add distinct documents from diffusion}
\ENDIF
\endgroup
\STATE $y \leftarrow \mathcal{M}(q, \mathcal{D}_{ret})$ \COMMENT{Generate final answer}
\RETURN $y$
\end{algorithmic}
\end{algorithm*}

\begingroup

\section{Further Experiments}
\label{sec:further_experiments}

\subsection{Statistical Significance Testing}
\label{sec:statistical_significance}

We compare G$^3$RAG with HippoRAG2 using Nv-embed-v2 on 1,000 paired evaluation queries per dataset. We apply a two-sided paired randomization test to the difference in mean F1 and estimate 95\% confidence intervals using paired bootstrap resampling of queries. The intervals characterize variability across evaluation queries. As shown in Table \ref{tab:statistical_significance}, G$^3$RAG achieves significantly higher F1 on MuSiQue and HotpotQA at the $0.05$ significance level. On 2WikiMultiHopQA, its F1 is 1.37 points lower, but the difference is not statistically significant ($p=0.26371$).

\begin{table}[t]
    \centering
    
    \caption{Statistical significance of F1 differences using Nv-embed-v2. $\Delta$ denotes G$^3$RAG minus HippoRAG2 in F1 points. Confidence intervals are for $\Delta$, and significance is assessed at $p<0.05$.}
    \label{tab:statistical_significance}
    \small
    \resizebox{\textwidth}{!}{%
    \begin{tabular}{l cccccc}
        \toprule
        Dataset & G$^3$RAG F1 & HippoRAG2 F1 & $\Delta$ & 95\% CI & $p$-value & Conclusion \\
        \midrule
        MuSiQue & 52.28 & 47.60 & $+4.68$ & $[2.35,7.02]$ & 0.00018 & Significant \\
        2WikiMultiHopQA & 69.20 & 70.57 & $-1.37$ & $[-3.76,1.02]$ & 0.26371 & Not significant \\
        HotpotQA & 78.00 & 75.30 & $+2.70$ & $[1.24,4.18]$ & 0.00034 & Significant \\
        \bottomrule
    \end{tabular}%
    }
\end{table}

\subsection{Effect of Seed Node Filtering}
\label{sec:seed_filtering}

We examine the role of $Filter_{\text{llm}}$ in G$^3$RAG and HippoRAG2. Table \ref{tab:seed_filtering} shows that removing this component reduces average F1 by 5.96 points for G$^3$RAG and 6.79 points for HippoRAG2, indicating that both methods benefit from filtering their initial nodes. Without filtering, G$^3$RAG exceeds HippoRAG2 by 2.00 points in average EM and 2.83 points in average F1.

\begin{table}[t]
    \centering
    
    \caption{Effect of seed node filtering using Nv-embed-v2. Each ablation removes $Filter_{\text{llm}}$ from the corresponding method. Bold indicates the best result in each column.}
    \label{tab:seed_filtering}
    \small
    \resizebox{0.80\textwidth}{!}{%
    \begin{tabular}{l cccccccc}
        \toprule
        \multirow{2}{*}{Method} & \multicolumn{2}{c}{MuSiQue} & \multicolumn{2}{c}{2Wiki} & \multicolumn{2}{c}{HotpotQA} & \multicolumn{2}{c}{Average} \\
        \cmidrule(lr){2-3} \cmidrule(lr){4-5} \cmidrule(lr){6-7} \cmidrule(lr){8-9}
        & EM & F1 & EM & F1 & EM & F1 & EM & F1 \\
        \midrule
        HippoRAG2 & 36.70 & 47.60 & \textbf{63.80} & \textbf{70.57} & \textbf{62.40} & 75.30 & 54.30 & 64.49 \\
        \quad w/o $Filter_{\text{llm}}$ & 28.90 & 39.40 & 57.30 & 63.00 & 57.90 & 70.69 & 48.03 & 57.70 \\
        \midrule
        G$^3$RAG & \textbf{39.10} & \textbf{52.28} & 62.30 & 69.20 & 62.20 & \textbf{78.00} & \textbf{54.53} & \textbf{66.49} \\
        \quad w/o $Filter_{\text{llm}}$ & 33.70 & 45.22 & 57.30 & 63.03 & 59.10 & 73.33 & 50.03 & 60.53 \\
        \bottomrule
    \end{tabular}%
    }
\end{table}

This dependence on filtering follows from the distinction between graph construction and seed selection. The offline geometric graph encodes complementary document relationships independently of the query, but cannot determine whether an initial node is relevant to that query. Diffusion from an incorrect seed can therefore propagate retrieval errors. Consistent with this limitation, G$^3$RAG without filtering yields a slightly lower average F1 than Naive RAG (60.53 versus 60.62 in Table \ref{tab:ablation}).

The ablations in Table \ref{tab:ablation} also distinguish the contribution of filtering from that of geometric graph retrieval. Adding $Filter_{\text{llm}}$ to Naive RAG raises average F1 from 60.62 to 63.39. With filtering enabled in both methods, G$^3$RAG further improves average F1 to 66.49 and average EM from 52.07 to 54.53, with gains in both metrics on all three datasets. These results indicate that geometric graph retrieval contributes additional gains beyond filtering, while relying on reliable semantic anchors for effective diffusion.

\subsection{Comparison with MMR and DPP}
\label{sec:mmr_dpp_comparison}

\begin{wraptable}{r}{0.60\textwidth}
    \vspace{-\intextsep}
    \setlength{\abovecaptionskip}{0pt}
    \setlength{\belowcaptionskip}{7pt}
    \centering
    
    \caption{Comparison with MMR and DPP using Nv-embed-v2 under identical candidate budgets and model settings. Bold indicates the best result in each column.}
    \label{tab:mmr_dpp_comparison}
    \small
    \setlength{\tabcolsep}{2pt}
    \resizebox{\linewidth}{!}{%
    \begin{tabular}{l cccccccc}
        \toprule
        \multirow{2}{*}{Method} & \multicolumn{2}{c}{MuSiQue} & \multicolumn{2}{c}{2Wiki} & \multicolumn{2}{c}{HotpotQA} & \multicolumn{2}{c}{Average} \\
        \cmidrule(lr){2-3} \cmidrule(lr){4-5} \cmidrule(lr){6-7} \cmidrule(lr){8-9}
        & EM & F1 & EM & F1 & EM & F1 & EM & F1 \\
        \midrule
        Naive    & 34.20 & 46.21 & 56.10 & 60.41 & 59.10 & 75.25 & 49.80 & 60.62 \\
        MMR      & 26.90 & 40.13 & 47.80 & 51.42 & 52.00 & 65.87 & 42.23 & 52.47 \\
        DPP      & 34.90 & 47.19 & 56.40 & 60.80 & 58.90 & 74.79 & 50.07 & 60.93 \\
        \midrule
        G$^3$RAG & \textbf{39.10} & \textbf{52.28} & \textbf{62.30} & \textbf{69.20} & \textbf{62.20} & \textbf{78.00} & \textbf{54.53} & \textbf{66.49} \\
        \bottomrule
    \end{tabular}%
    }
\end{wraptable}

We compare G$^3$RAG with Maximal Marginal Relevance (MMR) and determinantal point processes (DPPs) under identical candidate budgets and model settings. As shown in Table \ref{tab:mmr_dpp_comparison}, DPP outperforms MMR but provides only modest improvements over Naive RAG in average EM and F1. G$^3$RAG exceeds DPP by 4.46 points in average EM and 5.56 points in average F1, with gains in both metrics across all three datasets. These results suggest that document diversification alone is insufficient for retrieving complementary evidence for multi-hop reasoning, supporting the use of relevance-conditioned novelty in graph construction and retrieval.

\subsection{Global versus Local Density Estimation}
\label{sec:density_scaling}

\begin{wraptable}{r}{0.56\textwidth}
    \vspace{-\intextsep}
    \setlength{\abovecaptionskip}{0pt}
    \setlength{\belowcaptionskip}{7pt}
    \centering
    
    \caption{Comparison of global and local density estimates using Nv-embed-v2. $\Delta$ denotes Global minus Local, computed before rounding.}
    \label{tab:density_scaling}
    \small
    \setlength{\tabcolsep}{3pt}
    \resizebox{\linewidth}{!}{%
    \begin{tabular}{l cccccc}
        \toprule
        \multirow{2}{*}{Dataset} & \multicolumn{2}{c}{Global} & \multicolumn{2}{c}{Local} & \multicolumn{2}{c}{$\Delta$} \\
        \cmidrule(lr){2-3} \cmidrule(lr){4-5} \cmidrule(lr){6-7}
        & EM & F1 & EM & F1 & EM & F1 \\
        \midrule
        MuSiQue  & \textbf{39.10} & \textbf{52.28} & 36.80 & 49.00 & +2.30 & +3.28 \\
        2Wiki    & \textbf{62.30} & \textbf{69.20} & 61.10 & 67.95 & +1.20 & +1.25 \\
        HotpotQA & \textbf{62.20} & \textbf{78.00} & 62.10 & 77.95 & +0.10 & +0.05 \\
        \midrule
        Pooled   & \textbf{54.53} & \textbf{66.49} & 53.33 & 64.97 & +1.20 & +1.53 \\
        \bottomrule
    \end{tabular}%
    }
\end{wraptable}

Motivated by prior findings on hubness and its mitigation through local and global similarity scaling \cite{radovanovic2010hubs,schnitzer2012local,conneau2017word}, we replace the default global density penalty with a local density penalty by estimating $\rho_j$ from $k_{\rho}=10$ nearest neighbors. The neighborhood size is fixed across datasets without tuning. Graph vertices and directed edges, the $[0.1,0.7]$ similarity truncation, embeddings, the geometric gain factor $x\sqrt{1-x^2}$, dataset-specific $\beta$, $Filter_{\text{llm}}$, $s=2$, $n=5$, and generation settings are unchanged.

As shown in Table \ref{tab:density_scaling}, the global density penalty achieves higher EM and F1 on all three datasets, with the largest gains on MuSiQue. On average, the global density penalty improves EM by 1.20 points and F1 by 1.53 points over the local alternative. These results suggest that the global density penalty better captures the similarity distribution across the entire retrieval corpus. In contrast, the local penalty can be dominated by highly similar candidates in the current neighborhood, limiting its effectiveness in suppressing redundancy.

\subsection{Seed Count under a Fixed Retrieval Budget}
\label{sec:seed_budget}

In the evaluated G$^3$RAG configuration, the final retrieval budget of $n=5$ comprises $s=2$ retained seed documents and three additional documents selected through diffusion. Setting $s=n=5$ leaves no budget for additional evidence, reducing the final context to the filtered seed documents. The default HippoRAG2 configuration uses $s=5$, but both methods return five documents to the generator. Thus, the different seed counts reflect their retrieval configurations under the same final document budget.

The sensitivity analysis in Figure \ref{fig:beta_sensitivity}(d)--(f), conducted with $n=10$, shows that performance initially increases and then decreases as $s$ grows. Too few seeds can provide unreliable semantic anchors, whereas too many occupy retrieval slots that could otherwise accommodate novel evidence. This trade-off motivates allocating the retrieval budget between seed retention and diffusion.

\begin{table}[t]
    \centering
    
    \caption{Comparison of seed counts using Nv-embed-v2 with $Filter_{\text{llm}}$ enabled and a fixed retrieval budget of $n=5$. Bold indicates the best result in each column.}
    \label{tab:seed_budget}
    \small
    \resizebox{0.90\textwidth}{!}{%
    \begin{tabular}{l cccccccc}
        \toprule
        \multirow{2}{*}{Method} & \multicolumn{2}{c}{MuSiQue} & \multicolumn{2}{c}{2Wiki} & \multicolumn{2}{c}{HotpotQA} & \multicolumn{2}{c}{Average} \\
        \cmidrule(lr){2-3} \cmidrule(lr){4-5} \cmidrule(lr){6-7} \cmidrule(lr){8-9}
        & EM & F1 & EM & F1 & EM & F1 & EM & F1 \\
        \midrule
        Naive w/ $Filter_{\text{llm}}$ & 36.70 & 50.27 & 58.10 & 63.42 & 61.40 & 76.47 & 52.07 & 63.39 \\
        HippoRAG2 ($s=5$) & 36.70 & 47.60 & \textbf{63.80} & \textbf{70.57} & \textbf{62.40} & 75.30 & 54.30 & 64.49 \\
        HippoRAG2 ($s=2$) & 37.00 & 48.14 & 60.10 & 65.97 & 62.00 & 75.52 & 53.03 & 63.21 \\
        \midrule
        G$^3$RAG ($s=2$) & \textbf{39.10} & \textbf{52.28} & 62.30 & 69.20 & 62.20 & \textbf{78.00} & \textbf{54.53} & \textbf{66.49} \\
        \bottomrule
    \end{tabular}%
    }
\end{table}

Table \ref{tab:seed_budget} further evaluates HippoRAG2 with $s=2$. Relative to $s=5$, its average EM and F1 decrease by 1.27 and 1.28 points, respectively. Although both metrics improve on MuSiQue and F1 improves slightly on HotpotQA, performance declines on 2Wiki. With $s=2$ in both graph methods, G$^3$RAG achieves higher EM and F1 on all three datasets, indicating that its advantage persists when the seed count is matched.

\subsection{Scalability on a Synthetically Expanded Corpus}
\label{sec:synthetic_scalability}

The dense formulation in Algorithm \ref{alg:offline} performs all pairwise document comparisons and requires $O(N^2)$ time and storage. Here, zero-token construction refers to the absence of additional LLM calls and token consumption during graph construction, while embedding and offline indexing still incur computational costs. For scalability evaluation, we implement sparse graph construction using FAISS HNSW to retrieve a fixed number of approximate nearest-neighbor (ANN) candidates per document. Similarity filtering and density-penalized edge weighting are applied to these candidate edges, and the retained graph is stored in compressed sparse row (CSR) format. With at most $K$ retained neighbors per document, the graph stores $O(NK)$ edges without materializing the full $N \times N$ similarity matrix.

We synthetically expand MuSiQue to corpora containing 10,000 to 100,000 documents through replication. All replicated documents are re-embedded, and their embedding cost is included in the total indexing time. Table \ref{tab:synthetic_scalability} reports indexing time, storage requirements, query latency, and peak query memory.

\begin{table}[htbp]
    \centering
    
    \caption{Scalability of ANN-based sparse indexing on a synthetically expanded MuSiQue corpus. Total indexing time includes document embedding. Dense $\mathbf{W}$ storage is estimated for an $N \times N$ matrix in float64 format.}
    \label{tab:synthetic_scalability}
    \small
    \setlength{\tabcolsep}{4pt}
    \resizebox{\textwidth}{!}{%
    \begin{tabular}{r rrrrrrr}
        \toprule
        \shortstack{Documents\\$N$} & \shortstack{Total\\indexing\\(min)} & \shortstack{Graph\\construction\\(s)} & \shortstack{Index disk\\(GiB)} & \shortstack{CSR matrix\\(GiB)} & \shortstack{Dense $\mathbf{W}$\\estimate\\(GiB)} & \shortstack{Avg. query\\(ms)} & \shortstack{Peak query\\memory\\(GiB)} \\
        \midrule
        10,000  &  7.97 &  3.63 & 0.31 & 0.0048 &  0.75 & 85.25 & 0.28 \\
        20,000  & 15.93 &  9.33 & 0.63 & 0.0095 &  2.98 & 88.28 & 0.44 \\
        30,000  & 23.94 & 12.33 & 0.94 & 0.0141 &  6.71 & 81.20 & 0.60 \\
        40,000  & 31.82 & 16.44 & 1.25 & 0.0185 & 11.92 & 83.38 & 0.76 \\
        50,000  & 39.92 & 20.74 & 1.56 & 0.0228 & 18.63 & 80.37 & 0.91 \\
        60,000  & 47.70 & 23.32 & 1.87 & 0.0269 & 26.82 & 80.31 & 1.07 \\
        70,000  & 55.52 & 27.60 & 2.19 & 0.0310 & 36.51 & 81.09 & 1.22 \\
        80,000  & 63.55 & 31.13 & 2.50 & 0.0348 & 47.68 & 79.53 & 1.38 \\
        90,000  & 71.30 & 36.03 & 2.81 & 0.0385 & 60.35 & 81.17 & 1.53 \\
        100,000 & 79.26 & 37.80 & 3.12 & 0.0421 & 74.51 & 82.85 & 1.69 \\
        \bottomrule
    \end{tabular}%
    }
\end{table}

Across the tested range, total indexing time increases approximately linearly from 7.97 to 79.26 minutes, while average query latency ranges from 79.53 to 88.28 ms. At 100,000 documents, the complete index occupies 3.12 GiB on disk, including a 0.0421 GiB CSR matrix, compared with an estimated 74.51 GiB for a dense float64 matrix. These measurements support the computational feasibility of the sparse implementation on the tested synthetic corpora. They do not establish retrieval accuracy or deployment performance on real corpora containing millions of documents. Evaluation on diverse large document collections and validation in practical engineering systems remain directions for future work.
\par
\endgroup

\section{Case Studies}
\label{sec:appendix_C}

\begin{figure*}[t]
    \centering
    \includegraphics[width=1.0\textwidth]{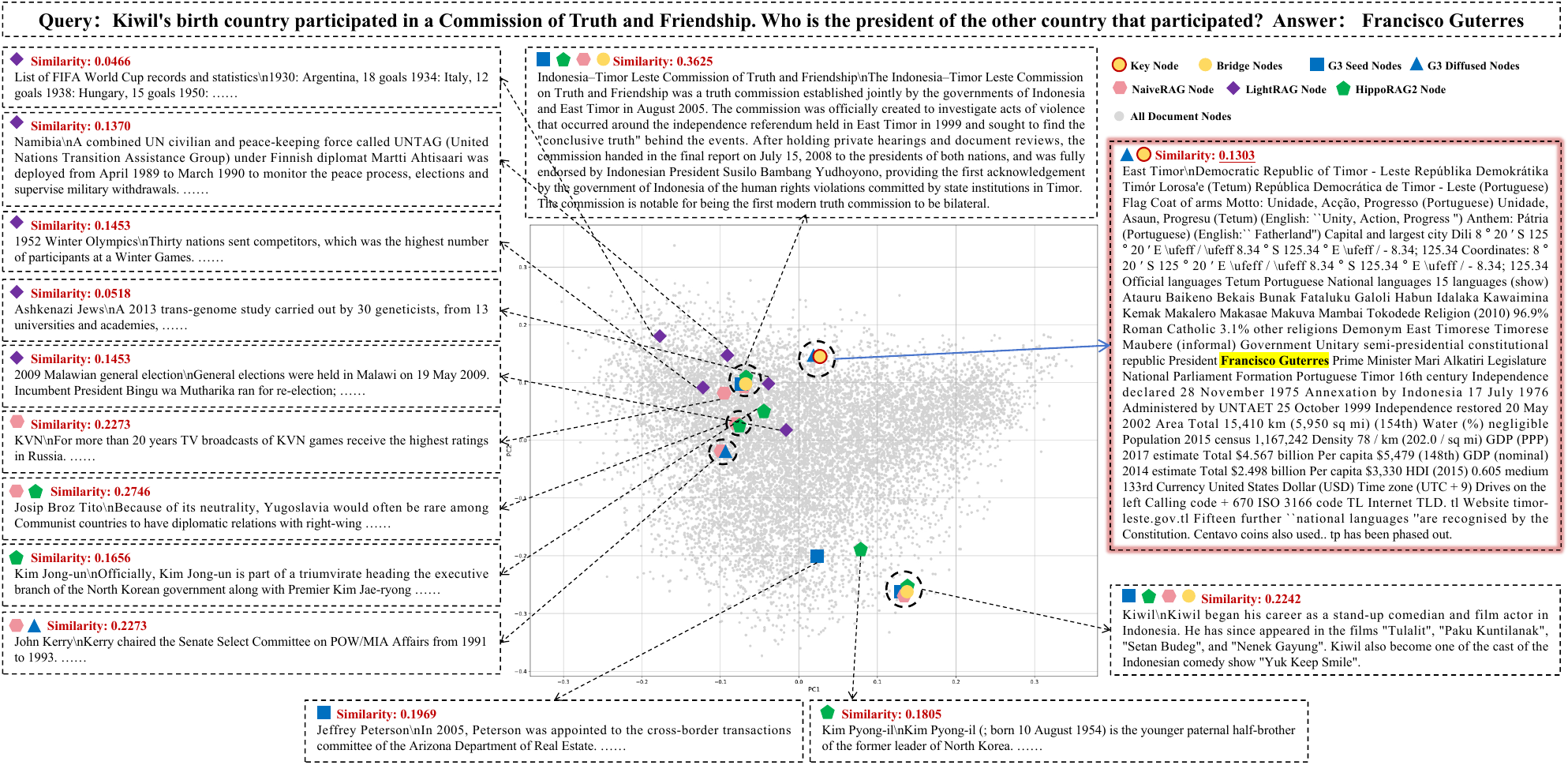}
    \caption{Qualitative comparison for example 1. The document enclosed in the red box is the most critical one serving as the source of the answer.}
    \label{fig:case_study1}
\end{figure*}

\begin{figure*}[t]
    \centering
    \includegraphics[width=1.0\textwidth]{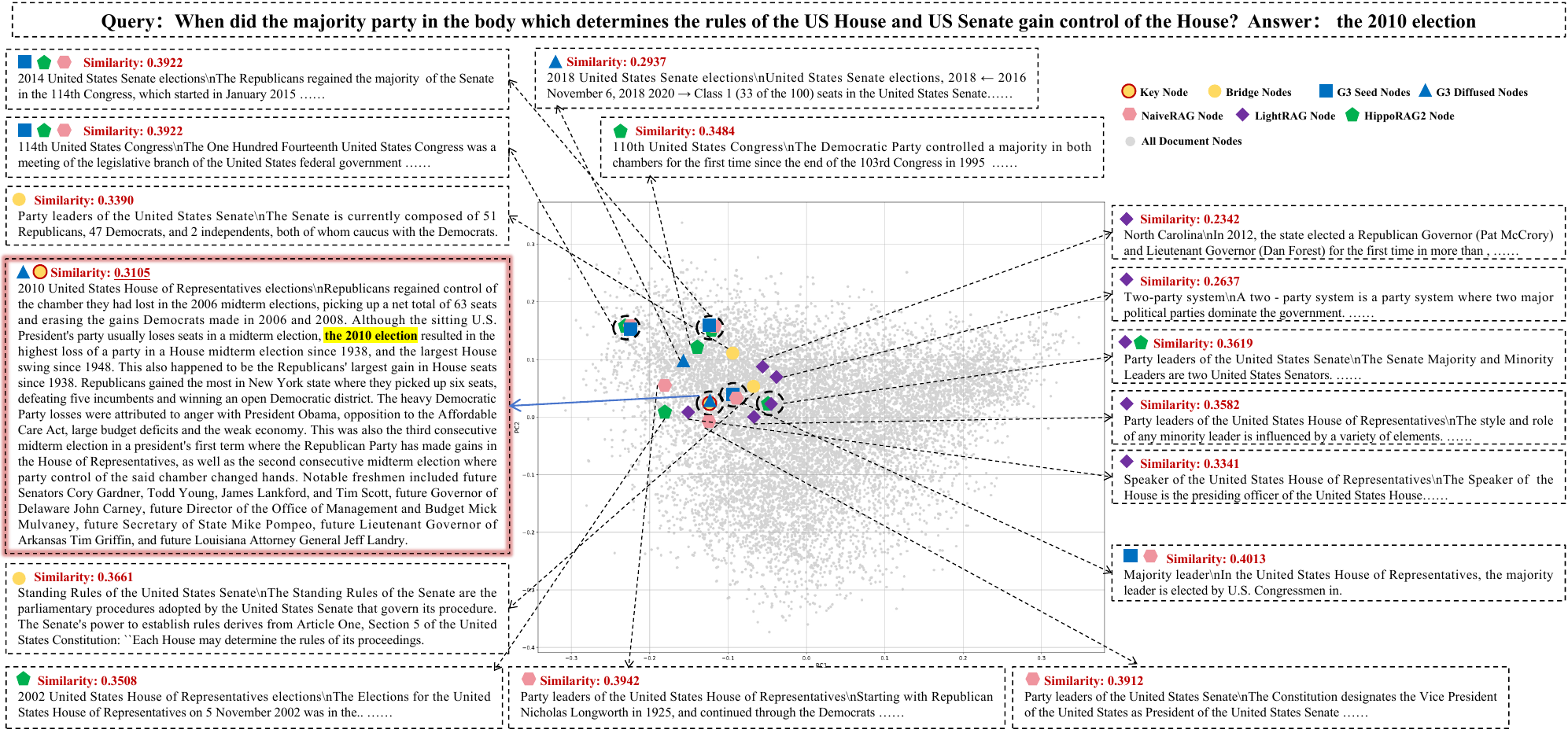}
    \caption{Qualitative comparison for example 2.}
    \label{fig:case_study2}
\end{figure*}

\begin{figure*}[t]
    \centering
    \includegraphics[width=1.0\textwidth]{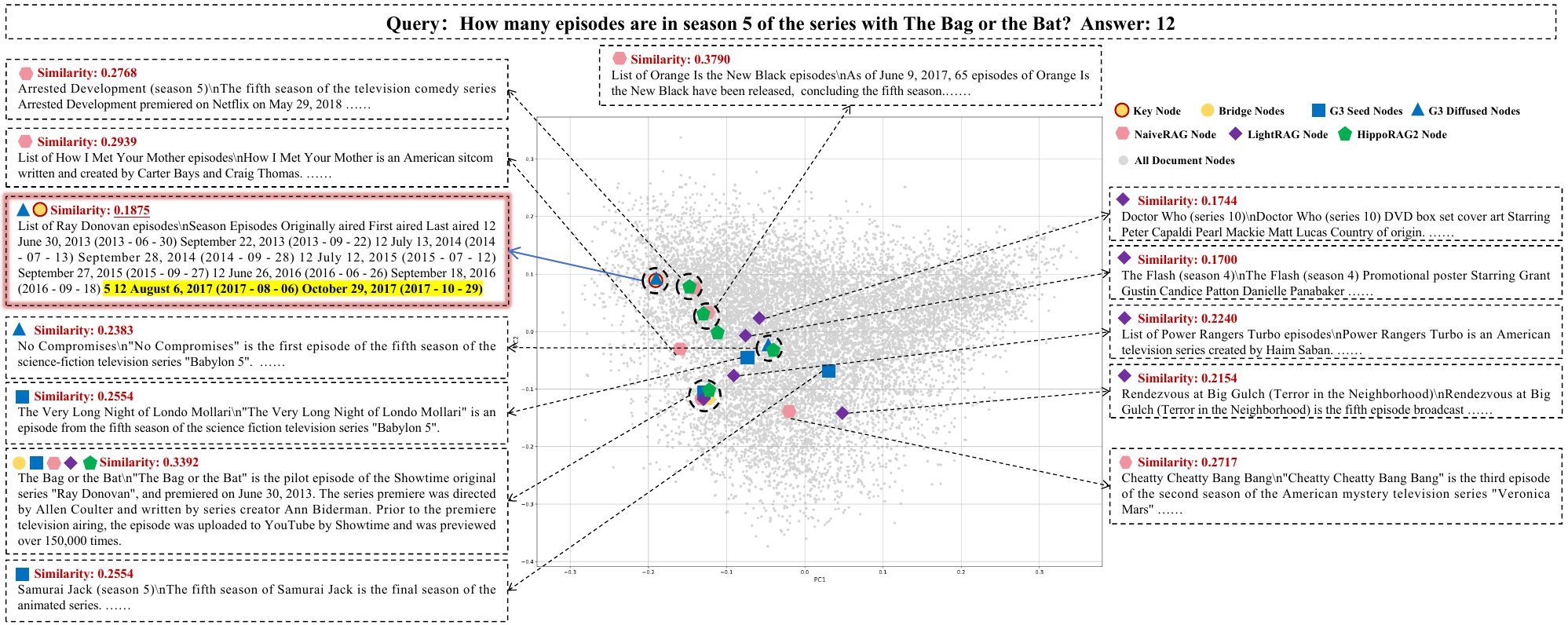}
    \caption{Qualitative comparison for example 3.}
    \label{fig:case_study3}
\end{figure*}

The preceding experiments demonstrate G$^3$RAG's significant advantage in discovering novel evidence documents on the highly challenging MuSiQue dataset. Consequently, we conducted instance-level visualizations of the retrieval behaviors of NaiveRAG, LightRAG, HippoRAG2, and G$^3$RAG on this dataset. Figures \ref{fig:case_study1} to \ref{fig:case_study3} visualize the retrieval behaviors of NaiveRAG, HippoRAG2, LightRAG, and G$^3$RAG across the entire document space. In these figures, except for bridge and key nodes that retain their complete textual content, all other nodes display only partial text. Detailed analyses are as follows:

As shown in Figure \ref{fig:case_study1}, the case study involving the query about "Kiwil's birth country" and the "Commission of Truth and Friendship"  effectively demonstrates the topological advantages of our proposed G$^3$RAG framework over existing retrieval baselines. Both Naive retrieval and HippoRAG succumb to the relevance trap, retrieving highly homogeneous context that fails to complete the necessary multi-hop reasoning chain. Although they successfully identify initial bridge documents like the "Commission of Truth and Friendship" and "Kiwil" , their rigid similarity constraints force the remaining retrieval window to be populated by topically adjacent but logically useless nodes, such as "Josip Broz Tito" or "Kim Jong-un", completely missing the terminal answer. Conversely, LightRAG suffers from severe semantic drift by over-exploring novelty, retrieving documents with extremely low similarity that entirely deviate from the core topic, such as the "1952 Winter Olympics" or "FIFA World Cup" statistics. In stark contrast, G$^3$RAG elegantly resolves this dichotomy through controlled geometric diffusion. By firmly anchoring the semantic trajectory with the high-relevance "Commission" and "Kiwil" seed nodes, G$^3$RAG leverages its pre-calculated orthogonal gain graph to channel the probability flow toward structurally significant, long-tail evidence. This controlled expansion allows G$^3$RAG to successfully retrieve the terminal document "East Timor" , which contains the crucial answer regarding President Francisco Guterres, proving that our method can accurately navigate from known premises to novel, low-similarity conclusions without triggering the exogenous noise seen in unconstrained exploration.

In the second case study regarding the historical timeline of congressional majority control (as shown in Figure \ref{fig:case_study2}), we observe that both Naive and HippoRAG predominantly fetch documents with high lexical overlap related to congressional leadership, such as the "2014 United States Senate elections" or the "114th United States Congress". Their retrieved sets form a dense cluster around these specific, yet temporally incorrect, historical events, entirely missing the actual answer document ("2010 United States House of Representatives elections" ) due to its lower raw similarity score of 0.3105. LightRAG exhibits a contrasting phenomenon characterized by broad topical dispersion; it retrieves generalized concepts like the "Two-party system" and localized events such as "North Carolina" state politics. These documents have exceedingly low similarity scores (0.2637 and 0.2342, respectively)  and do not contribute to the specific temporal reasoning required. In the G$^3$RAG framework, the retrieval log shows a two-stage operational phenomenon: it first anchors on the high-overlap documents (e.g., the 114th Congress) as initial seeds , and subsequently diffuses through the geometric gain graph to activate the target 2010 election document. This demonstrates how the diffusion process navigates from highly similar initial premises to a specific, lower-similarity terminal node within the same semantic trajectory.

In the third case study concerning the episode count for a specific television season (as shown in Figure \ref{fig:case_study3}), we observe a classic instance of lexical confinement. The query targets the fifth season of the series associated with the episode "The Bag or the Bat". Both Naive and HippoRAG successfully retrieve the initial bridge document, "The Bag or the Bat", which possesses a similarity score of 0.3392. However, their remaining retrieval windows form a dense cluster around documents with high keyword overlap (e.g., "season 5" and "episodes") but entirely unrelated television contexts, such as "List of Orange Is the New Black episodes" (0.3790) and "Arrested Development (season 5)" (0.2768). Consequently, they fail to reach the actual answer document, "List of Ray Donovan episodes", due to its significantly lower raw similarity score of 0.1875. LightRAG displays a phenomenon of broad topical dispersion. While it captures the bridge document , the rest of its retrieved set consists of disconnected series with low similarity scores, such as "List of Power Rangers Turbo episodes" (0.2240) and "The Flash (season 4)" (0.1700), which do not contribute to resolving the query. Conversely, the G$^3$RAG framework demonstrates a distinct two-stage operational phenomenon. It first anchors on the highly relevant bridge document "The Bag or the Bat"  as part of its initial seeds. Subsequently, it diffuses through the predefined graph to activate the specific terminal document, "List of Ray Donovan episodes". This illustrates how the geometric diffusion process successfully navigates from a localized semantic anchor to the exact, lower-similarity terminal node required to complete the multi-hop reasoning chain.

In summary, the qualitative case studies clearly demonstrate the distinct advantages of the G$^3$RAG framework over existing retrieval baselines. Built upon a \textbf{zero-token} offline graph construction, G$^3$RAG first anchors on highly relevant seed nodes and subsequently performs controlled topological diffusion across its unique geometric gain graph. This mechanism effectively avoids the "similarity trap" that confines methods like NaiveRAG and HippoRAG2 to locally homogenized regions. Furthermore, it prevents the severe semantic drift observed in LightRAG's unconstrained exploration. Ultimately, G$^3$RAG establishes a robust and precise retrieval trajectory, seamlessly crossing semantic gaps to efficiently capture multi-hop, high-information-gain documents that are crucial for complex reasoning.

\section{Derivation of the Optimal Theoretical Upper Bound}
\label{appendix:derivation}

In Section \ref{sec:methodology}, we defined the directed geometric information gain (edge weight) between two document representations as the product of their relevance and novelty. For any vector similarity $x = \cos\theta$ where $x \in [0, 1]$, the objective function is given by:
\begin{equation}
    g(x) = x \cdot \sqrt{1 - x^2} \label{eq:appendix_g}
\end{equation}

The primary objective of this derivation is to determine the theoretical upper bound of the similarity $x$ required to achieve the optimal theoretical geometric gain, providing a mathematical basis for our threshold selection. We present two equivalent derivations: an algebraic derivative approach and a geometric trigonometric approach. Both derivations rigorously demonstrate that the absolute mathematical maximum occurs at $x \approx 0.707$. Rather than claiming a strict empirical optimum for real-world retrieval tasks, we use this theoretical peak as a guide to strategically establish the upper limit of our empirical filtering interval at $0.7$, while setting a lower bound of $0.1$ to proactively mitigate severe semantic drift.

\paragraph{Derivation 1: Algebraic Derivative Analysis}
To find the global maximum of the continuous function $g(x)$ defined on the interval $x \in [0, 1]$, we compute its first derivative with respect to $x$ using the product rule and the chain rule:
\begin{align}
    g'(x) &= \frac{d}{dx} \left( x \cdot (1 - x^2)^{\frac{1}{2}} \right) \nonumber \\
          &= 1 \cdot (1 - x^2)^{\frac{1}{2}} + x \cdot \frac{1}{2}(1 - x^2)^{-\frac{1}{2}} \cdot (-2x) \nonumber \\
          &= \sqrt{1 - x^2} - \frac{x^2}{\sqrt{1 - x^2}} \label{eq:derivative}
\end{align}

To find the critical points, we set the first derivative to zero ($g'(x) = 0$):
\begin{align}
    \sqrt{1 - x^2} - \frac{x^2}{\sqrt{1 - x^2}} &= 0 \nonumber \\
    1 - x^2 &= x^2 \nonumber \\
    x^2 &= \frac{1}{2} \label{eq:critical}
\end{align}

Since $x$ represents the cosine similarity in the positive orthant ($x \ge 0$), we discard the negative root. Thus, the critical point is $x^* = \frac{\sqrt{2}}{2} \approx 0.707$. Evaluating the boundary conditions yields $g(0) = 0$ and $g(1) = 0$. Since $g(x) > 0$ for all $x \in (0, 1)$, the stationary point $x^*$ strictly guarantees the global maximum, yielding a maximum theoretical edge weight of $0.5$. The increase in $g(x)$ up to $x^*$ and its subsequent decrease make $x^*$ a theoretical upper bound for the region of increasing gain, providing a reference for selecting the maximum similarity threshold.

\paragraph{Derivation 2: Trigonometric Identity Perspective}
The elegance of this theoretical upper bound can be more intuitively observed through trigonometric identities. Recall the geometric definitions $x = \cos\theta$ and $\sqrt{1-x^2} = \sin\theta$. Equation \ref{eq:appendix_g} can be directly rewritten as a function of the semantic angle $\theta$:
\begin{equation}
    g(\theta) = \cos\theta \cdot \sin\theta = \frac{1}{2}\sin(2\theta)
\end{equation}

Given that the semantic angle $\theta \in [0, \frac{\pi}{2}]$ for non-negative similarities, the function $\sin(2\theta)$ achieves its strict maximum value of $1$ when $2\theta = \frac{\pi}{2}$, yielding $\theta^* = 45^{\circ}$. Mapping this optimal semantic angle back to the cosine similarity space yields the exact same stationary point:
\begin{equation}
    x^* = \cos(45^{\circ}) = \frac{\sqrt{2}}{2} \approx 0.707
\end{equation}

While both derivations rigorously establish $x^* \approx 0.707$ as the maximizer of $g(x)$ in a continuous vector space, it is crucial to clarify its role in the discrete, highly irregular manifold of real-world document embeddings. This result does not imply that document pairs with similarity $x^*$ provide better retrieval evidence than those at other similarity values. Our core motivation in proposing geometric gain edges is not to pursue a specific absolute value, but to establish a \textit{directional gradient}. The function $x \sqrt{1-x^2}$ acts as a continuous topological regularizer: it dynamically penalizes trajectories that collapse into semantic redundancy ($x \to 1$) or diverge into unpredictable semantic drift ($x \to 0$). The empirically optimal similarity threshold is difficult to determine because the embedding space representations vary across different models. However, by introducing a directional constraint, we can prevent the two extreme cases, allowing the graph to connect documents that exhibit both relevance and novelty. Therefore, guided by the theoretical bound  as a reference for threshold selection rather than a guarantee of empirical optimality, we set the upper limit of $x$ at $0.7$ (just below the $0.707$ peak) and enforce a lower-bound truncation at $0.1$ to proactively filter out off-topic documents.

\begingroup

\subsection{Alternative Gain Forms}
\label{sec:equivalent_gain_forms}

The geometric gain admits the equivalent expressions $g(x)=x\sqrt{1-x^2}=\cos\theta\sin\theta=\frac{1}{2}\sin(2\theta)$, where $x=\cos\theta$. Relevance and angular novelty are complementary functions of the same angle, and their product favors edges that exhibit both properties. Although $g(x)$ is increasing on the retained interval $[0.1,0.7]$ and thus preserves the ordering of cosine similarities, this does not imply equivalence in weighted diffusion.

Let $x_{ij}$ denote cosine similarity, $m_{ij}$ indicate whether an edge is retained, and $q_j=\exp(-\beta\rho_j)$ denote its target density penalty. For a node with outgoing edges, a gain function $f$ induces the transition probabilities
\begin{equation}
    T_{ij}^{(f)} = \frac{m_{ij} f(x_{ij}) q_j}{\sum_k m_{ik} f(x_{ik}) q_k}.
    \label{eq:gain_form_transition}
\end{equation}
Multiplying all outgoing weights of a node by the same positive constant leaves these probabilities unchanged. In contrast, $g(x)/x=\sqrt{1-x^2}$ varies across neighbors and decreases with similarity. Geometric gain therefore shifts probability toward less similar but still relevant neighbors compared with cosine weighting. For two neighbors with similarities $(0.7,0.4)$ and equal density penalties, normalized cosine weights are $(0.636,0.364)$, whereas geometric gain gives $(0.577,0.423)$. Such redistribution can change the final document ranking when diffusion scores are aggregated across seeds.

\begin{wraptable}{r}{0.56\textwidth}
    \vspace{-\intextsep}
    \setlength{\abovecaptionskip}{0pt}
    \setlength{\belowcaptionskip}{7pt}
    \centering
    
    \caption{Comparison of gain forms using Nv-embed-v2. All variants share the density penalty. Scores are F1, with the best result in each column in bold.}
    \label{tab:gain_forms}
    \small
    \setlength{\tabcolsep}{3pt}
    \resizebox{\linewidth}{!}{%
    \begin{tabular}{l cccc}
        \toprule
        Gain form & MuSiQue & 2Wiki & HotpotQA & Average \\
        \midrule
        $x^{0.5}$ & 49.09 & 68.83 & 77.69 & 65.20 \\
        $x^{0.8}$ & 48.50 & 68.72 & 77.52 & 64.91 \\
        $x^{0.9}$ & 49.22 & \textbf{69.67} & 77.70 & 65.53 \\
        $x$ (cosine) & 49.68 & 68.07 & 77.02 & 64.92 \\
        \midrule
        $x\sqrt{1-x^2}$ & \textbf{52.28} & 69.20 & \textbf{78.00} & \textbf{66.49} \\
        \bottomrule
    \end{tabular}%
    }
\end{wraptable}

To assess whether generic concave transformations provide comparable gains, we evaluate $w_{ij}^{(\gamma)}=x_{ij}^{\gamma}\exp(-\beta\rho_j)$ with $\gamma\in\{0.5,0.8,0.9\}$ alongside cosine and geometric weighting. Graph topology, the $[0.1,0.7]$ truncation, embeddings, density penalties, $Filter_{\text{llm}}$, $s=2$, $n=5$, and diffusion and generation settings are identical. The two filtered seed documents for each query are fixed across all variants. Given the limited evaluation samples, we report all three exponents without selecting one on a held-out validation set.

Table \ref{tab:gain_forms} shows that $\gamma=0.9$ achieves the highest average F1 among the tested concave controls, improving over cosine weighting by 0.61 points. Geometric gain improves over cosine by 1.57 points and over the strongest concave control by 0.96 points on average, although $x^{0.9}$ performs best on 2Wiki. These results indicate that generic concave calibration accounts for part of the observed improvement.

These results show that the tested alternative gain forms do not match the average F1 of the proposed geometric gain, despite preserving the ordering of cosine similarities. The choice of $\cos\theta$ and $\sin\theta$ is motivated by semantic relevance and novelty, respectively. The former measures directional consistency, while the latter models non-redundancy through the orthogonal component. Their product encodes the need for evidence that is both relevant and complementary, grounding the design in the retrieval objective rather than mathematical equivalence alone.
\par
\endgroup

\end{document}